\documentclass[journal, twocolumn]{IEEEtran}
\usepackage{tabularx}
\pdfoutput=1
\usepackage{url}
\usepackage[T1]{fontenc}
\usepackage{algorithmic}
\usepackage{comment}
\usepackage{placeins}
\usepackage{mathrsfs,amsmath}
\usepackage{graphicx}
\usepackage{amsmath,amssymb,amsfonts}
\usepackage{array}
\usepackage{textcomp}
\usepackage[font=small]{caption}
\usepackage{paralist}
\usepackage{lettrine}
\usepackage{cite}
\usepackage{xcolor}
\usepackage{cancel}
\usepackage{float}
\usepackage{booktabs}
\usepackage{subfig}
\usepackage{stfloats}
\usepackage{url}
\usepackage{lipsum}
\usepackage{enumitem}
\usepackage{mathtools}
\usepackage{cuted}
\usepackage{makecell}
\usepackage{acronym}
\usepackage{multicol}
\usepackage{multirow}
\providecommand{\keywords}[1]{\textbf{\textit{Index terms---}} #1}
\usepackage{glossaries-extra}
\setabbreviationstyle[acronym]{long-short}
\newacronym{TRP}{TRP}{total radiated power}
\newacronym{mmW}{mmW}{millimeter-wave}
\newacronym{LOS}{LOS}{line-of-sight}
\newacronym{5G}{5G}{fifth-generation}
\newacronym{B5G}{B5G}{beyond 5G}
\newacronym{IMT}{IMT}{international mobile telecommunications}

\newacronym{EIRP}{EIRP}{equivalent isotropically radiated power}
\newacronym{CF}{CF}{characteristic function}
\newacronym{TDD}{TDD}{time division duplex}
\newacronym{PDF}{PDF}{probability density function}
\newacronym{CDF}{CDF}{cumulative distribution function}
\newacronym{UPA}{UPA}{uniform planar array}
\newacronym{ULA}{ULA}{uniform linear array}
\newacronym{AoD}{AoD}{angle of departure}
\newacronym{PA}{PA}{power amplifier}
\newacronym{RFC}{RFC}{RF chain}
\newacronym{PSD}{PSD}{power spectral density}
\newacronym{U6G}{U6G}{upper 6GHz}
\newacronym{GSMI}{GSMI}{geometry-based stochastic model of the interference}
\newacronym{SMI}{SMI}{stochastic model of the interference}
\newacronym{THz}{THz}{terahertz}
\newacronym{UE}{UE}{user equipment}
\newacronym{SATFP}{SATFP}{satellite footprint}
\newacronym{EM}{EM}{electromagnetic}
\newacronym{SAT}{SAT}{satellite}
\newacronym{INR}{INR}{interference to noise ratio}
\newacronym{BS}{BS}{base station}
\newacronym{ITU}{ITU}{international telecommunications union}
\usepackage[ruled,vlined]{algorithm2e}
\usepackage{authblk}

\newcommand{\RN}[1]{%
  \textup{\uppercase\expandafter{\romannumeral#1}}%
}

\def\BibTeX{{\rm B\kern-.05em{\sc i\kern-.025em b}\kern-.08em
    T\kern-.1667em\lower.7ex\hbox{E}\kern-.125emX}}

\title{IMT to Satellite Stochastic Interference \\ Modeling and Coexistence Analysis of \\Upper 6 GHz Band Service}

\author{Reza~Aghazadeh~Ayoubi, Dario~Tagliaferri, Filippo~Morandi, Luca~Rinaldi, \\\vspace{-0.2cm}Laura~Resteghini, Christian Mazzucco and Umberto~Spagnolini
\thanks{R.A.\, Ayoubi, D.\ Tagliaferri, F.\,Morandi, L.\,Rinaldi and U.\ Spagnolini are with the  Department of Electronics, Information and Bioengineering (DEIB) of Politecnico di Milano, 20133 Milan, Italy  (e-mail: [reza.aghazadeh, dario.tagliaferri, filippo.morandi, luca.rinaldi, umberto.spagnolini]@polimi.it)}
\thanks{L. Resteghini and C. Mazzucco are with Huawei Technologies Italia S.r.l., Segrate, 20054 Italy. (e-mail: [laura.resteghini, christian.mazzucco]@huawei.com)}
\thanks{U.\ Spagnolini is Huawei Industry Chair at Politecnico di Milano}}
\usepackage[numbers,sort&compress]{natbib}

\begin{document}

\maketitle

\begin{abstract} 
The surging capacity demands of 5G networks and the limited coverage distance of high frequencies like millimeter-wave (mmW) and sub-terahertz (THz) bands have led to consider the upper 6GHz (U6G) spectrum for radio access. However, due to the presence of the existing satellite (SAT) services in these bands, it is crucial to evaluate the impact of the interference of terrestrial U6G stations to SAT systems. A comprehensive study on the aggregated U6G-to-SAT interference is still missing in the literature. In this paper, we propose a stochastic model of interference (SMI) to evaluate the U6G-to-SAT interference, including the statistical characterization of array gain and clutter-loss and considering different \textcolor{black}{interference modes}. Furthermore, we propose an approximate geometrical-based stochastic model of interference (GSMI) as an alternative method to SMI when the clutter-loss distribution is unavailable. Our results indicate that given the typical international
mobile telecommunication (IMT) parameters, the aggregated interference power is well below the relevant protection criterion, and we prove numerically that the GSMI method overestimates the aggregated interference power with only 2dB compared to the SMI method. 
\end{abstract}
\section{Introduction}
The coexistence of \gls{SAT} communications with \gls{5G} and \gls{B5G} \gls{BS} operating in the \gls{U6G} frequency is an arising issue due to the growing interest in 
new bands to increase capacity in densely populated areas \cite{5G_6GHz,coleagoWP,US_federal,AnalysisMason}. Studies demonstrate that the usage of the upper mid-band is necessary to fulfill the requirements of the downlink of 5G \cite{6GHz_Spc3}. At the same time, around $40 \%$ of the benefit foreseen for \gls{5G} mid-bands will not be exploited in the absence of a new mid-bands spectrum assignment \cite{6GHz_Spc2}. These additional frequencies provide large bandwidth, in excess of 100 MHz, while characterized by a smaller path loss compared to \gls{mmW} \gls{5G} \cite{Considerations_6GHz}. The deployment of new \gls{U6G} systems might affect the operation of \gls{SAT}s already in place that use these frequency bands in uplink \cite{Considerations_6GHz}, such as C-band (4-8 GHz) and X-band (8-12 GHz) \cite{NasaPaper}. Even if the emission of a single \gls{BS} serving all \gls{UE} has a negligible impact on \gls{SAT}, the aggregation of the interference from a large number of \gls{BS}s in a large area (e.g., the \gls{SATFP} on Earth) might be harmful. In \cite{Guarnieri}, the effect of the interference on geosynchronous synthetic aperture radars has been studied in the context of remote sensing in the C-band. However, the sources of interference are different w.r.t the \gls{IMT}. 

Currently, there are no comprehensive studies regarding the statistical analysis of aggregated interference from \gls{BS}s, observed by the \gls{SAT}s in the \gls{U6G} bands. The coexistence analysis between \gls{IMT}-2020 and \gls{SAT} systems has been widely studied (see, e.g., \cite{mmWaveModellling0,mmWaveModellling1,mmWaveModellling4,mmWaveModellling5,mmWaveModellling7}) mostly for \gls{mmW} bands (24.25 and 86 GHz), and in the context of relevant agendas (see e.g., \cite{WRC15} or \gls{ITU} agenda WRC-23  item 1.2). The previous works usually target the aggregated interference in the \gls{SATFP} considering only the direct \gls{BS}-\gls{SAT} path.
Moreover, they all consider the same propagation model, valid for frequencies above 10 GHz (see \cite{iturp2108} for further details). 
Therefore, it is clear that existing interference modeling approaches cannot be readily extended to the \gls{U6G} spectrum, since \textit{(i)} the propagation model is not appropriate for frequencies $<10$ GHz, and \textit{(ii)} different \textcolor{black}{interference modes} are typically neglected. The term \textcolor{black}{interference mode is herein used for any propagation that ends up toward the \gls{SAT}, including the direct \gls{BS}-\gls{SAT} path or reflections (from the ground or buildings towards the \gls{SAT}).} 

The problem of interference estimation in communication systems involves the modeling of both the \gls{U6G} devices (deployment, functioning, antennas, etc.) and propagation for the involved frequencies and environments. 
Several guidelines to evaluate the compatibility between terrestrial and space stations are provided in the literature. The work in \cite{iturm2101} presents a methodology for modeling IMT-Advanced, namely fourth-generation (4G), and IMT-2020 (5G), networks, and systems for general coexistence studies. It details the simulation setup, including the modeling of network topology and antenna arrays. The methodology is based on the characteristics of \gls{IMT}-advanced systems \cite{iturm2292}. 

Besides system modeling, it is necessary to determine a suitable propagation model for earth-space interference evaluation \cite{iturp619}, including all the relevant phenomena such as clutter loss,
which is an additional loss with respect to the path-loss, created by the diffraction, reflection, or scattering of the buildings and vegetation in the vicinity of the \gls{BS}s.
An empirical model for the \gls{CDF} of the clutter loss is reported in \cite{iturp2108}, for earth-space links working above 10 GHz. This latter model can be used when the geometry of the scenario is not known. In contrast, when prior information on the environment is available, e.g., statistical characterization of the geometrical features, the stochastic model in \cite{iturp2402} might be applied, provided that appropriate modifications are made to extend its validity below 10 GHz.

The main contribution of this paper is the development of a stochastic method that can be used to evaluate the aggregate interference at the \gls{SAT} from U6G terrestrial \gls{BS}s. The proposed method is general since it does not constrain the analysis to any specific scenario. The detailed contributions are listed in the following: 
\begin{itemize}
    \item We develop an \gls{SMI} towards a \gls{SAT} from a set of micro and macro \gls{BS}s, based on a stochastic description of the \gls{BS} array gain and clutter loss, calculated according to the geometrical distribution of a given region. We use a \gls{CF}-based approach, to efficiently aggregate all the interference power, from different types of \gls{BS}s, when serving both indoor and outdoor UEs, to ultimately yield a methodology for estimating the aggregated interference power from the \gls{SATFP}. 
    
    \item We propose a \gls{GSMI} method to estimate the aggregated interference at the \gls{SAT} when no clutter-loss statistics are available. The GMSI method leverages the environment's geometrical statistics. 
     
    \item We provide numerical examples of the interference \gls{CDF} for \gls{U6G} service with the SMI and GSMI methods. Our results demonstrate that, given typical parameter values, the aggregated interference power from \gls{U6G} is well below the \gls{INR} protection criterion, while it is relevant only for extreme values of the employed parameters. We show that the GSMI results overestimate the interference power density by only $2$ dB with respect to SMI results, on aggregate for the \gls{SATFP}.
    
\end{itemize}

\textit{Organization}: The remainder of the paper is organized as follows: in Sec.~\ref{Sec:SysModel} we present the system model. In Sec.~\ref{Sec:InterfAnalysis} and Sec.~\ref{Sec:Aggregate}, the methodology for modeling the interference from a single \gls{BS} and from \gls{BS}s in a large region are presented, respectively. The distribution of the array gain and clutter-loss are discussed in Sec.~\ref{Sec:Array} and Sec.~\ref{Sec:CL}. In Sec.~\ref{Sec:geom_stat}, the process of extracting the geometrical statistics is discussed. The numerical results are in Sec.~\ref{Sec:NumericalResults} and the paper is concluded in Sec.~\ref{Sec:Conclusion}.
\section{System Model}\label{Sec:SysModel}
\begin{figure}[!tb]
\hspace{0cm}
     \includegraphics[width = 0.99 \columnwidth]{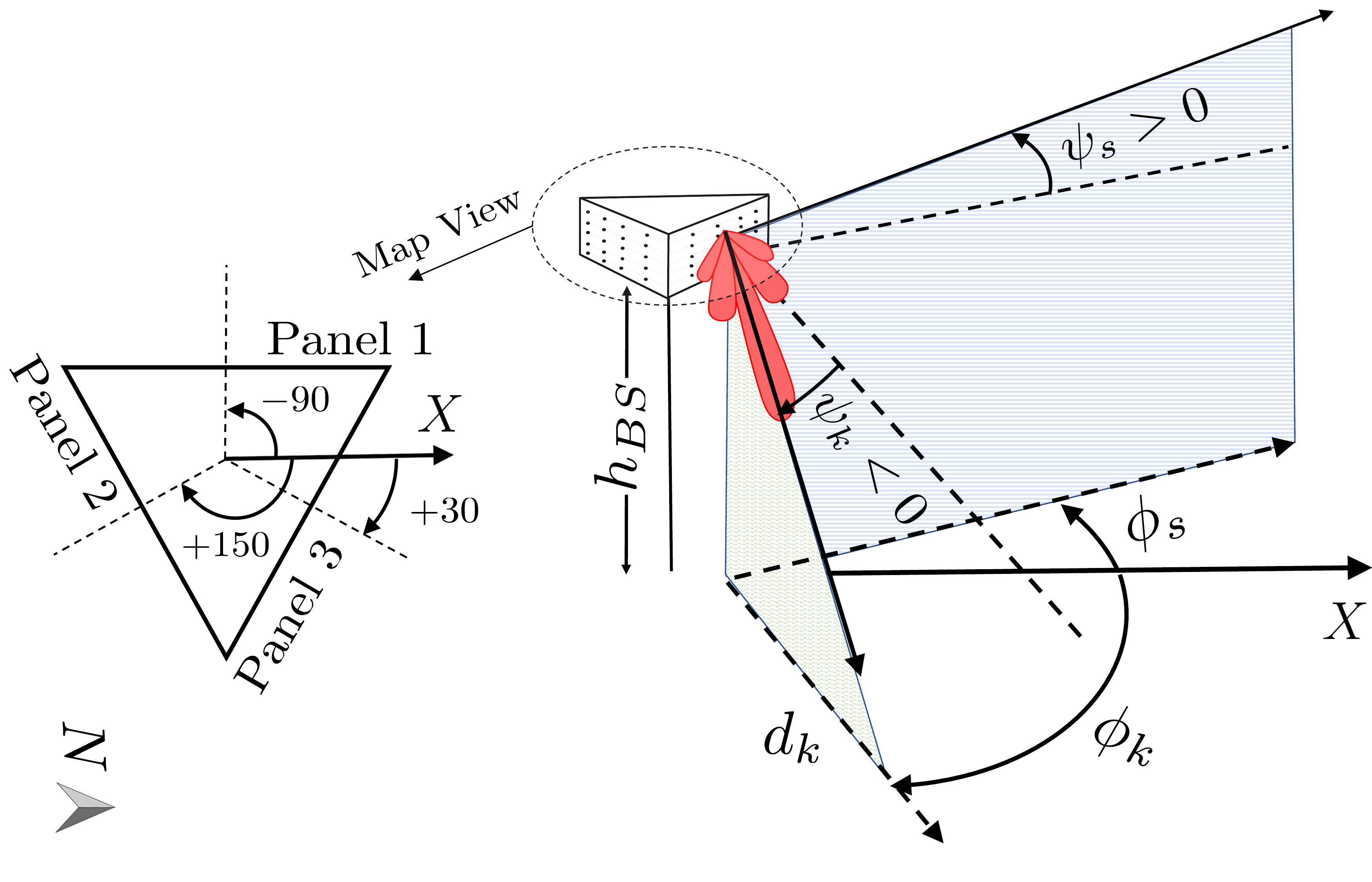}
    \caption{Tri-sectoral \gls{BS} serving a UE, while unwillingly generating at the \gls{SAT}. }\label{fig:BS-serv}
\end{figure}
\begin{figure}[!t]
\centering
    \includegraphics[height = 5cm]{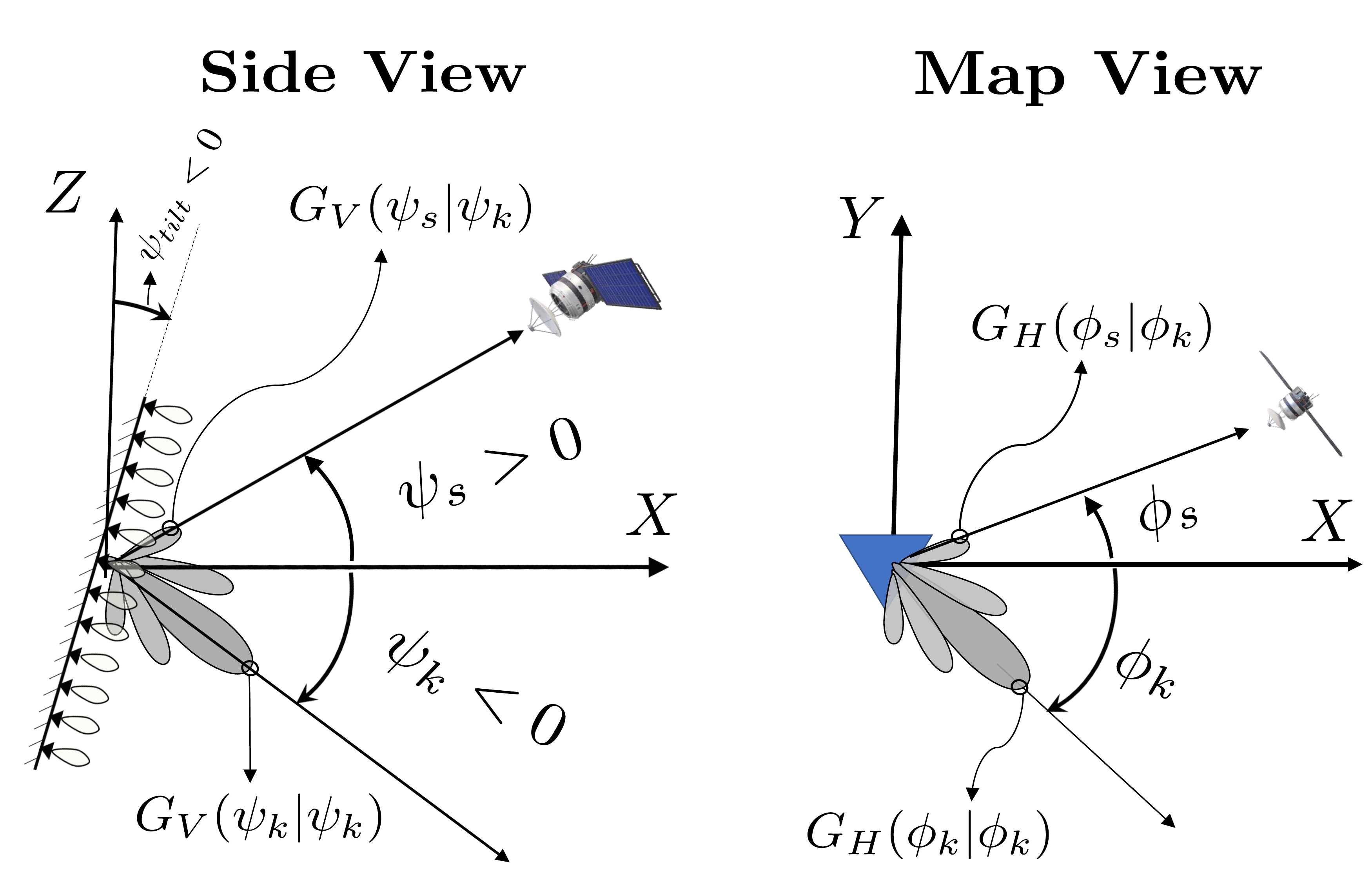}
    \caption{ Map view  and side view of the tri-sectoral panels. \label{fig:Array}}
\end{figure}

Modeling the aggregated interference to a \gls{SAT} from a set of \gls{BS}s requires both geometrical and propagation considerations. Let us consider the scenario in Figure~\ref{fig:BS-serv}, where a single \gls{BS} at height $h_{BS}$ is causing interference to the \gls{SAT} while serving a single-antenna ground UE. The coordinate system is such that an arbitrary set of angles $\boldsymbol{\vartheta}=(\psi,\phi)$, consists of azimuth angle $-180<\phi\leq 180$ deg, defined as clockwise positive from North, and elevation angle defined as $-90<\psi\leq 90$ deg relative to the ground plane, located at the \gls{BS} height. The \gls{SAT}, given a longitude and latitude of observation, is identified by the \gls{AoD} $\boldsymbol{\vartheta}_s = (\psi_s,\phi_s)$. In this setting, $\psi<0$ correspond to any \textcolor{black}{interference mode} that first bounces on the ground. The same coordinate system is used for the served UE at $\boldsymbol{\vartheta}_k = (\psi_k,\phi_i)$. A set of stochastic parameters $\boldsymbol{\Theta}$ characterizes the environment, \gls{UE}, and the \gls{BS} distributions, including: \textit{(i)} inter-building distance $d$; \textit{(ii)} \gls{BS}-building distance; \textit{(iii)} the \gls{BS} height $h_{BS}$; \textit{(iv)} buildings height $h$; \textit{(v)} UEs height $h_{UE}$. The signal received by the $k$-th UE from a single \gls{BS} is
\begin{equation}
    y_{k} = \sqrt{P_{T} G_a(\boldsymbol{\vartheta}_k|\boldsymbol{\vartheta}_k,\boldsymbol{\Theta})\alpha_k}\, x_k + w_k,\label{eq:UESignal}
\end{equation}
where $x_k$ is the Tx signal,  $G_a(\boldsymbol{\vartheta}_k|\boldsymbol{\vartheta}_k,\boldsymbol{\Theta})$ is the array gain toward the UE of interest, when the \gls{BS} array is designed to points toward $\boldsymbol{\vartheta}_k$ (Fig~.\ref{fig:Array}), $P_{T}$ is the  Tx power and $\alpha_k$ the path-loss for distance $d_{k}$, including any shadowing and fading and ${w_k}$ is the noise amplitude. For a given position and height of the \gls{BS}, the signal \eqref{eq:UESignal} toward the UE, generates interference at SAT. This is originated from either the direct path (i.e. at angle $\boldsymbol{\vartheta}_s$), and/or from other \textcolor{black}{interference modes}. For example, the signal $x_k$ might be reflected by ground/buildings toward the \gls{SAT}, or it might be diffracted by vegetation/building edges. Let $\boldsymbol{\vartheta}_{\ell}$ denote the \gls{AoD} of the rays in $\ell$-th propagation mode, which is a function of $\boldsymbol{\vartheta}_s$ (e.g., for direct \gls{BS}-SAT propagation mode, it is $\boldsymbol{\vartheta}_{\ell} = \boldsymbol{\vartheta}_{s}$). The interfering signal received by the \gls{SAT}, when the $q$-th \gls{BS} is serving the $k$-th UE ($\ell$-th \textcolor{black}{interference mode}) is
\begin{equation}
    \iota_{q,k}^{\ell}(\boldsymbol{\Theta}) = \sqrt{G_a(\boldsymbol{\vartheta}_{\ell}|\boldsymbol{\vartheta}_k,\boldsymbol{\Theta})\frac{P_{T}}{A_c(\boldsymbol{\vartheta}_{\ell} \lvert\boldsymbol{\Theta})}\alpha_s}\,\, x_k + w_s,\label{eq:InterfModel}
\end{equation}
where $G_a(\boldsymbol{\vartheta}_{\ell}|\boldsymbol{\vartheta}_k,\boldsymbol{\Theta})$ is the BS array gain toward $\boldsymbol{\vartheta}_{\ell}$ when it is designed to points to $\boldsymbol{\vartheta}_k$, $A_c(\boldsymbol{\vartheta}_{\ell}\lvert\boldsymbol{\Theta})$ is the clutter loss between the \gls{BS} and SAT, ${w_s}$ is the additive white Gaussian noise with \gls{PSD} $N_0$ over bandwidth $B$, while $\alpha_s$ consists of all the phenomena above the terrain as
\begin{equation}
    \alpha_s = \frac{G_s}{A_{s} A_{pol}},
\end{equation}
where $G_s$, $A_{s}$, and $A_{pol}$ are respectively the \gls{SAT} antenna gain, the free space path loss, and the loss due to polarization mismatch.
The dependence of the array gain and clutter loss on the set of geometrical parameters $\boldsymbol{\Theta}$, and the \textcolor{black}{interference modes} is detailed in Sec.~\ref{Sec:Array} and Sec.~\ref{Sec:CL}. Note that the beam spread loss, which is the loss caused by refractive effects of the atmosphere, is neglected since it is only relevant for very small elevation angles \cite{iturp6195}, and atmospheric gases absorption is typically neglected around 6 GHz \cite{NasaPaper,Atm_absorption}.

Although the largest part of the aggregated interference comes from the direct \gls{BS}-\gls{SAT} path, all the other \textcolor{black}{interference modes} cannot be neglected, otherwise, the interference is underestimated.
The aggregated interference power caused by $Q$ total \gls{BS}s each serving $K$ possible UEs, through $L$ possible \textcolor{black}{interference modes}, is
\begin{equation}
    I(\boldsymbol{\Theta}) =\sum_{q=1}^{Q}\sum_{k=1}^{K}\sum_{\ell=1}^{L} I^{\ell}_{q,k}(\boldsymbol{\Theta}),\label{eq:AggAll}
\end{equation}
where $I^{\ell}_{q,k}(\boldsymbol{\Theta}) = (\iota^{\ell}_{q,k}(\boldsymbol{\Theta}))^2$ is the single interference power contribution.
In practice, the number of served UEs $K$ is not deterministic, and the aggregation over all the served UEs can be replaced by modeling the transmit power $P_{T}$ with an appropriate \gls{PDF} and the \gls{BS} loading factor. Thus, herein the UE index $k$ in \eqref{eq:AggAll} will be dropped with the corresponding summation. 

\begin{figure}[htb!]
\centering
\includegraphics[scale = 0.35]{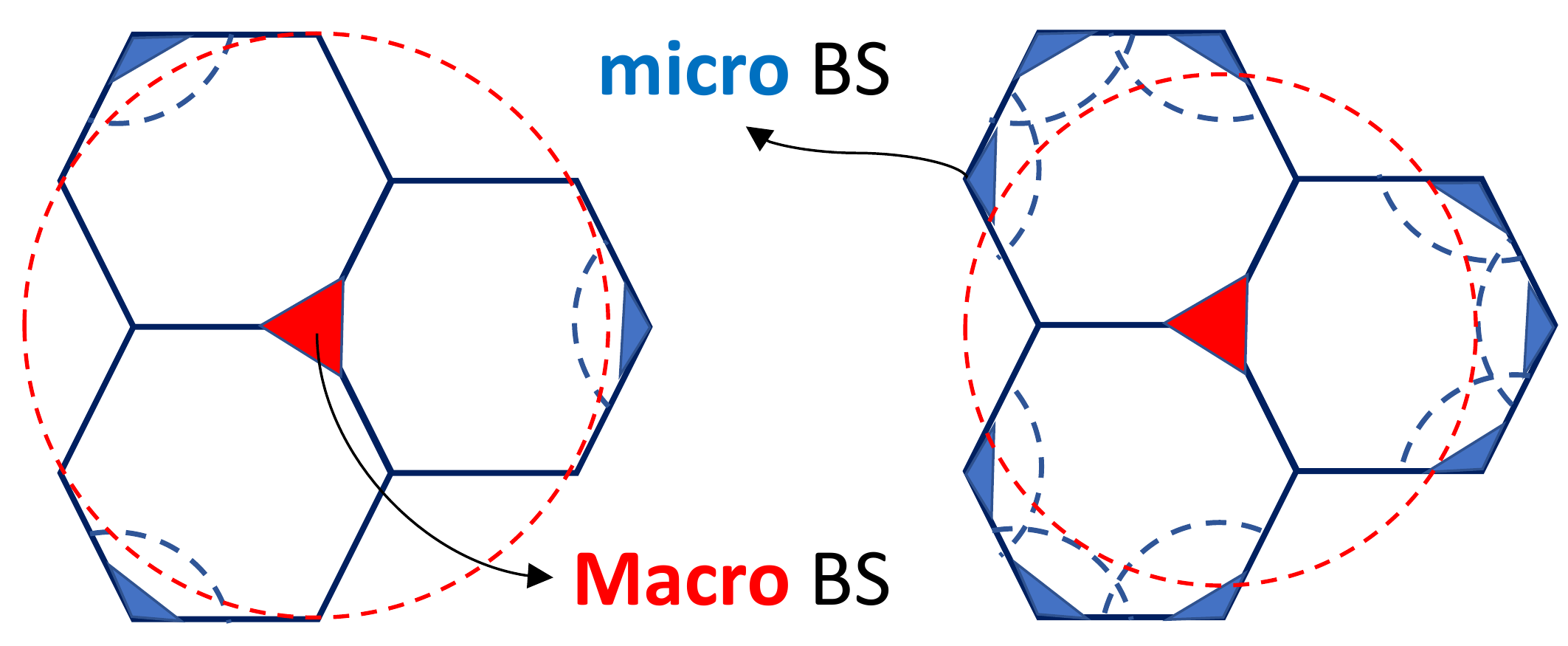}
\caption{Macro and micro \gls{BS} arrangement. Every cell is covered with one or multiple micro \gls{BS}s. Each micro \gls{BS} has a single sector array, that serves the edge UEs of the corresponding macro cell. }\label{fig:Cell_Shapes}
\end{figure}

\gls{BS}s can be modeled as transmitting with full power (On) or not transmitting at all (Off) \cite{iturm2101}, with a loading factor $\rho$ defined as the percentage of the \gls{BS}s that are randomly chosen as active. \textcolor{black}{Furthermore, each \gls{BS} transmit only a fraction of total time $F_T$, due to employing \gls{TDD}.} Each \gls{BS} is either a macro \gls{BS} or a micro \gls{BS}, as shown in Fig.~\ref{fig:Cell_Shapes}.
Macro \gls{BS}s employ larger array sizes, organized in three sectors to cover multiple cells, a higher transmitter power $P_T$, and a larger height compared to micro \gls{BS}. We assume, without any loss of generality, that macro \gls{BS}s are placed on top of the tallest building in each area \cite{iturm2101}, for coverage purposes. Differently, micro \gls{BS}s have a single sector, and they are characterized by a reduced Tx power and are mostly aimed at boosting coverage and capacity at cell edges. Thus, for mere modeling purposes, we assume the micro \gls{BS} is located on the ground, at the furthest distance from the macro one.
Considering a single \gls{BS}, its height from ground $h_{BS}$ as well as its position with respect to surrounding buildings can be regarded as random. This affects the modeling of the \textcolor{black}{interference modes} toward the \gls{SAT}, which can be evaluated in a probabilistic framework, using the geometrical statistics of the environment. 
\section{Statistical Model of Interference from a Single BS}\label{Sec:InterfAnalysis}

We propose a statistical framework to evaluate the interference at the \gls{SAT}, using both \gls{SMI} and \gls{GSMI}. In the \gls{SMI} method, all the possible \textcolor{black}{interference modes} (from every possible bounce of the rays) are considered to occur, and each propagation path is subject to a specific clutter loss, with a corresponding probability distribution. In the \gls{GSMI} method instead, the \textcolor{black}{interference modes} are limited to the significant ones, each of these occurring with a specific probability, while the clutter loss is not considered.
A main difference between the two methods is that the \gls{PDF} of the clutter loss in the SMI method is achieved by ray tracing. Instead, the GSMI makes use of stochastic geometry to approximate clutter loss.
\begin{figure}[tb!]
\centering
\includegraphics[width=1\columnwidth]{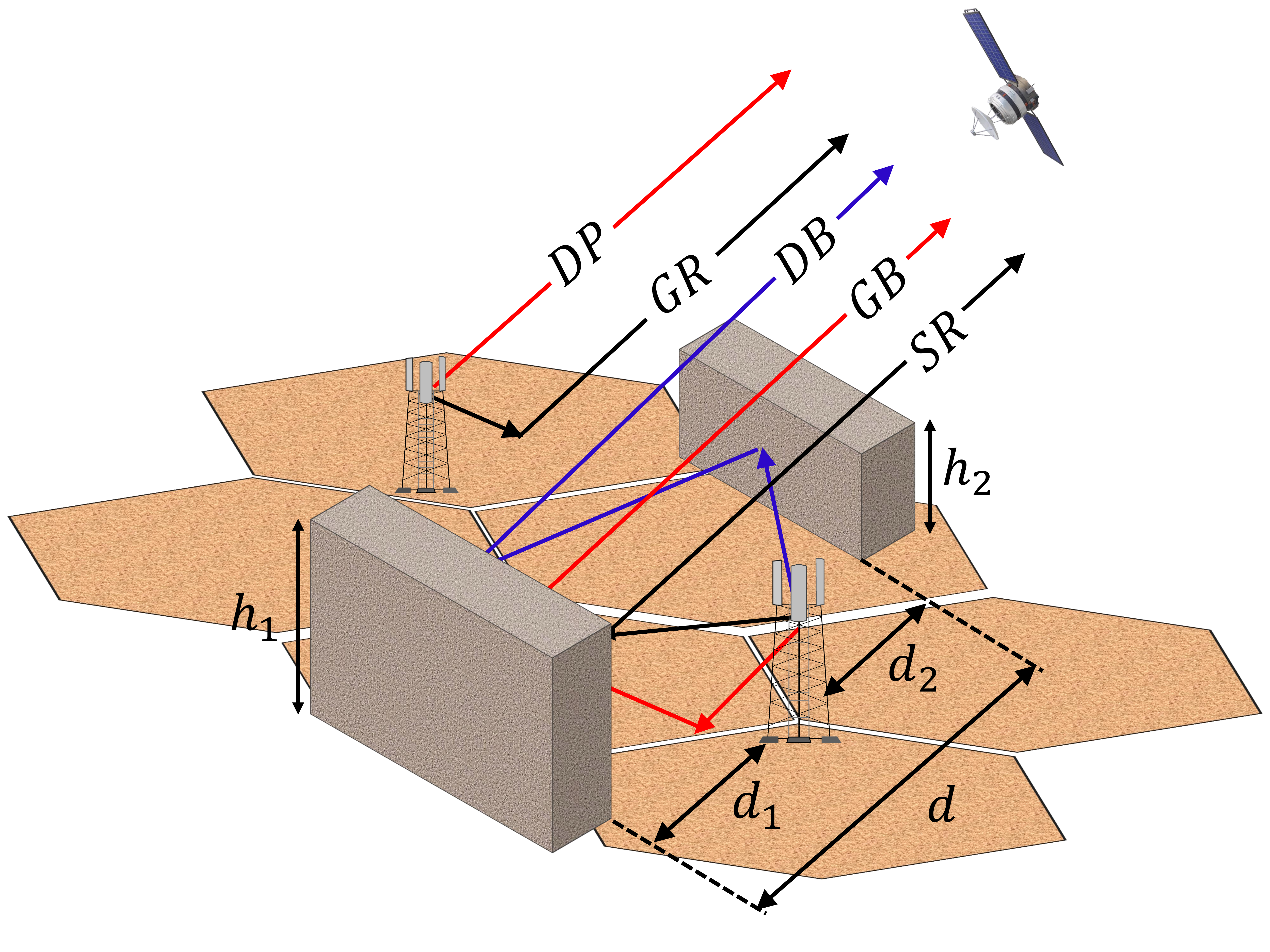}
\caption{\label{fig:scenario_1} \textcolor{black}{Interference modes} of GSMI method for a micro \gls{BS}}
\end{figure}
%
%
\subsection{Stochastic model of interference (SMI)}\label{Sec:SMI}
In the SMI method, the interference is evaluated by considering every possible \textcolor{black}{interference mode} that reaches the SAT with any number of bounces. The SMI method requires knowledge of the \gls{PDF} of both array gain and clutter-loss for every propagation mode. For the ${\ell}$-th propagation mode, the rays departing from the \gls{BS} with specific AoD $\boldsymbol{\vartheta}_{\ell}$ reach the \gls{SAT} experiencing a different array gain and clutter loss. Given the geometrical stochastic parameters $\boldsymbol{\Theta}$, the interference power at the SAT from the $q$-th \gls{BS} can be evaluated by adopting \eqref{eq:InterfModel} in dB scale as
\begin{align}
     \left[ I^{\ell}_{q}(\boldsymbol{\Theta}) \right] = & [P_{T}] + [G_a\left(\boldsymbol{\vartheta}_{\ell} \lvert \boldsymbol{\vartheta}_k,\boldsymbol{\Theta}\right)]+ [G_c(\boldsymbol{\vartheta}_{\ell}\lvert\boldsymbol{\Theta})] + [\alpha_s],\label{eq:interf_power}
\end{align}
where $[x]$ denotes the value of $x$ in dB scale, $G_c(\boldsymbol{\vartheta}_{\ell}|\boldsymbol{\Theta})=A^{-1}_c(\boldsymbol{\vartheta}_{\ell}|\boldsymbol{\Theta})$ is the clutter gain, inverse of the clutter loss defined in \eqref{eq:InterfModel}. The corresponding \gls{PDF} of $\left[ I^{\ell}_{q}(\boldsymbol{\Theta}) \right] $ is obtained as (\ref{eq:PDF_I}) (bottom of the page) by means of the \textit{logarithmic convolution} \cite{PDF_conversion}.
\begin{figure*}[!b]
\hrulefill
\begin{equation}
\begin{split}
    \mathcal{P}\left(\left[I^{\ell}_{q}\lvert\boldsymbol{\Theta})\right]\right) = \mathcal{P}\left([P_{T}]\right) * \mathcal{P}\left([G_a \lvert \boldsymbol{\vartheta}_{\ell},\boldsymbol{\Theta})]\right)*  \mathcal{P}\left([G_c\lvert \boldsymbol{\vartheta}_{\ell},\boldsymbol{\Theta}]\right) * \mathcal{P}\left([\alpha_s]\right)\label{eq:PDF_I}.
\end{split}
\end{equation}
\end{figure*}
The \gls{PDF} of the interference power in linear scale, i.e. $\mathcal{P}\left(I^{\ell}_{q}\lvert\boldsymbol{\Theta} \right)$, can be easily converted from dB scale as indicated in \cite{PDF_conversion}. Given the joint \gls{PDF} of the geometrical parameters $\boldsymbol{\Theta}$, $\mathcal{P}({\boldsymbol{\Theta}}\lvert\phi_s)$, we can write:
\begin{align}
    \mathcal{P}(I^{\ell}_{q}) =&{\mathbb{E}}_{\boldsymbol{\Theta}}\left[ \mathcal{P}(I^{\ell}_{q} \lvert \boldsymbol{\Theta})\right] = \int...\int \mathcal{P}(I^{\ell}_{q}\lvert\boldsymbol{\Theta})\mathcal{P}({\boldsymbol{\Theta}}\lvert\phi_s)d\boldsymbol{\Theta},\label{eq:decondition}
\end{align}
where $\mathbb{E}_z[x]$ is the expectation of x over $z$. Note that $\mathcal{P}({\boldsymbol{\Theta}}\lvert\phi_s)$ is conditioned to a given satellite azimuth $\phi_s$, as detailed in Sec.~\ref{Sec:geom_stat}. The analysis of the interference power is carried out using the \gls{CF} of the $\mathcal{P}(I^{\ell}_{q})$, defined hereafter as $\Phi^\ell_{q}(\omega) \triangleq \mathbb{E} \{ e^{i \omega I^{\ell}_{q}}\}$. The usage of the \gls{CF} is preferred in interference analysis \cite{UcylA,haenggi}, because it always exists when it is a function of a real-valued argument \cite{CF_EugenLukas}, and cumbersome convolution operations can be converted to simpler products. The aggregated interference from $q$-th \gls{BS} to SAT, $I_q$, is the independent summation (in linear scale \cite{powerSum, PowerSum2}) over the $L$ possible \textcolor{black}{interference modes}, whose \gls{CF} is achieved as 
\begin{equation}\label{eq:CF_AggScenarios}
    \Phi_{q}(\omega) = \prod_{\ell=1}^L {\Phi_q^\ell(\omega)}.
\end{equation}
This \gls{CF} is used in Sec.~\ref{Sec:Aggregate} for aggregation of interference coming from all \gls{BS}s in a given region.

\subsection{Geometry-based stochastic model of interference (GSMI)}\label{Sec:GSMI}
The GSMI method is an alternative to SMI whenever the \gls{PDF} of clutter gain $\mathcal{P}\left([G_c\lvert \boldsymbol{\vartheta}_{\ell},\boldsymbol{\Theta}]\right)$ in \eqref{eq:PDF_I} is not available. Typically, $\mathcal{P}\left([G_c\lvert \boldsymbol{\vartheta}_{\ell},\boldsymbol{\Theta}]\right)$ is obtained by means of exhaustive and computationally intensive ray-tracing simulations, which could be unavailable in some cases, especially over large areas such as the SATFP. With GSMI, we assume that the interference at the SAT comes from a limited set of \textcolor{black}{interference modes}. These are the following: \textit{(i)} direct path (DP); \textit{(ii)} single-building (SB) reflection; \textit{(iii)} double building (DB) reflection; \textit{(iv)} ground reflection (GR); \textit{(v)} ground and building (GB) reflection, while other reflections are neglected due to higher propagation losses. Herein, we denote the set of considered modes as ${\mathcal{M} = \{DP, SB,DB,GR,GB\}}$. The $\ell$-th \textcolor{black}{interference mode} $\ell \in \mathcal{M}$ can occur with a certain probability $\mathcal{P}_{\ell}(\boldsymbol{\Theta})$, that depends on the system parameters $\boldsymbol{\Theta}$ as well as on the \gls{BS} type. For instance, micro \gls{BS} can experience \textit{all} the interference modes, while for macro \gls{BS}, located on the rooftop, SB and DB, typically do not occur. Note that, one or more interference modes might occur simultaneously, and thus we have
\begin{equation}
    0\leq \sum^{|\mathcal{M}|}_{\ell=1}\mathcal{P}_{\ell}(\boldsymbol{\Theta})\leq |\mathcal{M}|.
\end{equation} 
Appendix \ref{Appendix:OccurrenceProb} reports the derivation of $\mathcal{P}_{\ell}(\boldsymbol{\Theta})$ and further information. 

Unlike the SMI method, where the possible \textcolor{black}{interference modes} $L$ is usually large, here the interference is limited to only 5 contributions (3 in case of rooftop BS). The average probability of occurrence of the $\ell$-th mode can be computed as
\begin{equation}
    \bar{\mathcal{P}}_{\ell}=\mathbb{E}_{\boldsymbol{\Theta}}\left[\mathcal{P}_{\ell}\left(\boldsymbol{\Theta}\right)\right].\label{eq:deconditionOcurrence}
\end{equation}
The interference power and its \gls{PDF} in each \textcolor{black}{interference mode} $\ell$ from $q$-th \gls{BS}, $\mathcal{P}(I^{\ell}_{q})$, is achieved with \eqref{eq:interf_power},
\eqref{eq:PDF_I} and \eqref{eq:decondition} by removing the clutter gain and its \gls{PDF}. However, it must be noted that, since every \textcolor{black}{interference mode} considered in the GSMI method has a specific occurrence probability, the PDF of the interference is conditioned to the occurrence of the corresponding $\ell$-th mode. 
Thus, this difference can be modeled by slightly modifying (\ref{eq:CF_AggScenarios}), yielding 
\begin{equation}
    \Phi_{q}(\omega) = \prod_{\ell=1}^{|\mathcal{M}|}\left( {\Phi^{\ell}_{q}(\omega)}\right)^{\bar{\mathcal{P}}_{\ell}}\label{eq:agg_GSMI}.
\end{equation}
This modification is justified in Sec.~\ref{Sec:Aggregate} when aggregating the interference coming from all \gls{BS}s in a given region. 

\section{Aggregation of Multiple BS}\label{Sec:Aggregate}

The interference power generated by a single \gls{BS} is then aggregated over multiple \gls{BS} such as over a city, or a large geographical region (e.g., the whole \gls{SATFP}). 

\subsection{Aggregation over a city}

The first aggregation step is to consider a whole area of a city $S$. The CF of the aggregated interference power at SAT from all the \gls{BS}s (either macro or micro) is computed as
\begin{equation}\label{eq:agg_SMI_GSMI}
  \Phi_{I}(\omega)=\begin{cases}
    \prod\limits_{\ell=1}^L \left({\Phi_q^\ell(\omega)}\right)^Q, & \text{SMI}.\\
    \prod\limits_{\ell=1}^{|\mathcal{M}|}\left( {\Phi^{\ell}_{q}(\omega)}\right)^{\bar{\mathcal{P}}_{\ell}Q}, & \text{GSMI}.
  \end{cases}
\end{equation}
where $Q = S\rho\lambda F_{T}$ is the effective average number of \gls{BS}s in the city, $\lambda$ is the density of the macro/micro \gls{BS}s, and $\rho$ is the \gls{BS} power loading factor based on the \gls{ITU} recommendation \cite{iturm2292} as the percentage $\rho$ of all the \gls{BS}s considered to be working at full power with maximum Tx power, \textcolor{black}{and $F_{T}$ is the \gls{BS} \gls{TDD} activity factor.} Model \eqref{eq:agg_SMI_GSMI} endorses that in the GSMI method, every \textcolor{black}{interference mode} occurs on average $\bar{\mathcal{P}}_{\ell}Q$ times, which is the rationale behind $\bar{\mathcal{P}}_{\ell}$ in \eqref{eq:agg_GSMI}.

The general aggregation rule \eqref{eq:agg_SMI_GSMI} can be specialized to derive the \gls{CF} in more specific cases, i.e., differentiating between different BS types (macro and micro) and UE locations (indoor vs. outdoor). The \gls{BS} type affects the interference mostly through the height $h_{BS}$, which changes from macro to micro and affects the \gls{PDF} of the array gain. Similar behavior is expected for the UE location (described by average UE height $h_{UE}$), as UEs located outdoor have $h_{UE}\approx 0$ while indoor UEs may have a much higher height from the ground. These assumptions affect the array gain and clutter loss. For example, let us consider the case of micro BSs, and assume that all the UEs are indoor. In this case, we have a constant BS height $h_{BS}$ m by assumption (see Sec.~\ref{Sec:SysModel}), and the \gls{UE} is bound to the building height $h$ as $0<h_{UE}< h$. The former term influences directly the clutter loss $A_c(\boldsymbol{\vartheta}_{\ell}\lvert\boldsymbol{\Theta})$ (see Sec.~\ref{Sec:CL}), while the second indirectly affects the array gain, defining a specific AoD toward the k-th UE, $\boldsymbol{\vartheta}_{k}$. We denote the aggregated interference under these assumptions as $I_{m,i}$ and its \gls{CF} with $  \Phi_{I}^{m,i}(\omega)$. The table \ref{tab:InterfCat}summarizes the four different interference contributions over an entire city. 
Let $0\leq\beta\leq1$ be the fraction of outdoor UEs and $1-\beta$ be the fraction of indoor \gls{UE}s. The overall \gls{CF} of the aggregated interference power is
\begin{align}\label{eq:agg_city}
    \Phi_{I}(\omega) \hspace{-0.1cm}=\hspace{-0.1cm} &\left(\Phi_{I}^{m,o}(\omega) \Phi_{I}^{M,o}(\omega)\right)^{\beta}
    \hspace{-0.2cm}\times\hspace{-0.1cm}\left(\Phi_{I}^{m,i}(\omega) \Phi_{I}^{M,i}(\omega)\right)^{1-\beta}.
\end{align}
Note that all of the micro and macro BSs coexist simultaneously, while a fraction of UEs is indoor/outdoor. The corresponding \gls{PDF} $\mathcal{P}(I)$ is computed as the inverse Fourier transform of the CF as
\begin{equation}
    \mathcal{P}(I)= \frac{1}{2\pi} \int_{-\infty}^{\infty}\Phi_{I}(\omega) e^{i\omega I}d\omega.
\end{equation}
In practice $\mathcal{P}(I)$ is evaluated using the Gil Pelaez theorem of inversion \cite{GP_Inversion}, which can be carried out numerically \cite{CF_Inv}.
\begin{table}[tb!]
\caption{Interference contributions}\label{tab:InterfCat}
  \centering
  {\renewcommand{\arraystretch}{1.4}
\begin{tabular}{|c|c|c|}
\hline 
\gls{BS} Type & UE Type & Notation   \tabularnewline
\hline 
\hline 
Micro \gls{BS}  
& Indoor
& $I_{m,i}$ and $  \Phi_{I}^{m,i}(\omega)$

\tabularnewline
\hline 
Micro \gls{BS}  
& Outdoor
& $I_{m,o}$ and $  \Phi_{I}^{m,o}(\omega)$
\tabularnewline
\hline

Macro \gls{BS}  
& Indoor
& $I_{M,i}$ and $  \Phi_{I}^{M,i}(\omega)$
\tabularnewline
\hline

Macro \gls{BS}  
& Outdoor
& $I_{M,o}$ and $  \Phi_{I}^{M,o}(\omega)$
\tabularnewline
\hline
\end{tabular}
}
\end{table}
\begin{figure}[tb!]
\centering
\subfloat[]{\includegraphics[scale=0.5]{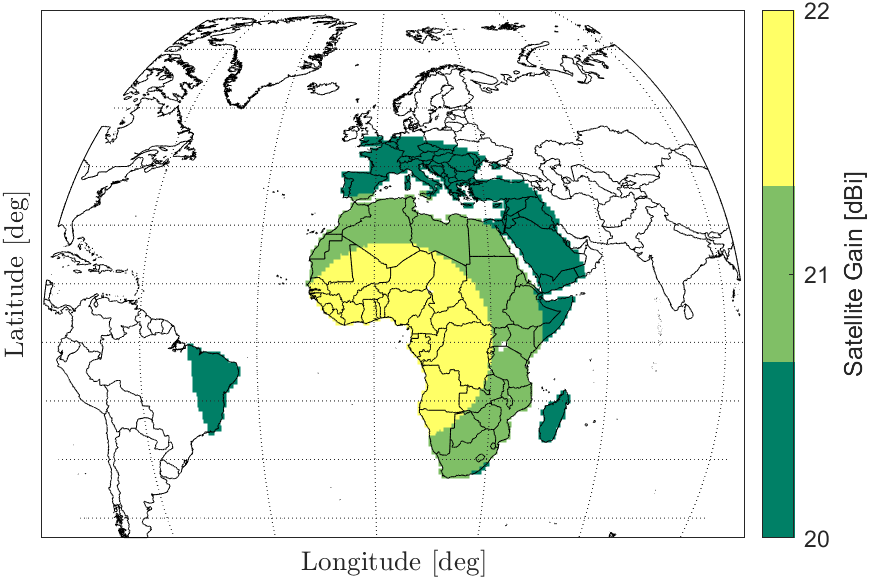}
  \label{fig:satgain} }

  \subfloat[]{ \includegraphics[scale=0.5]{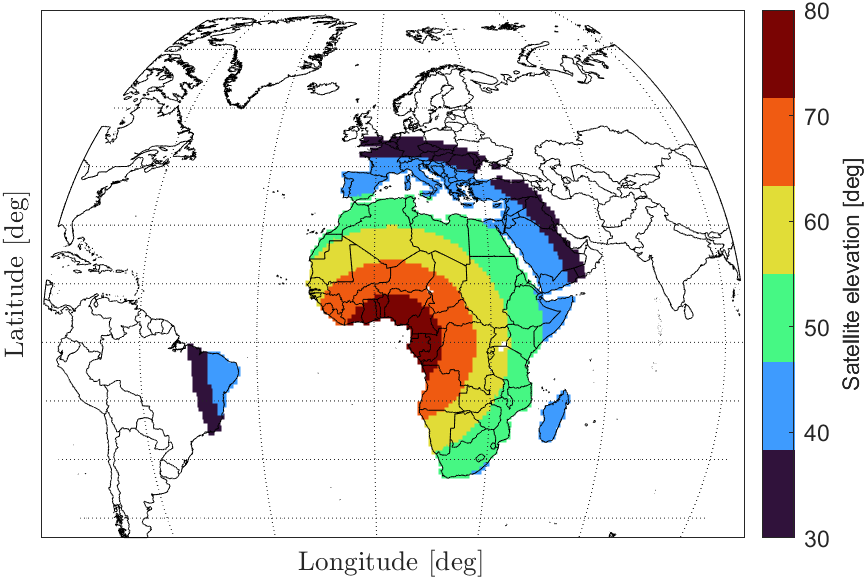}
  \label{fig:satelev} }

  \subfloat[]{\includegraphics[scale=0.5]{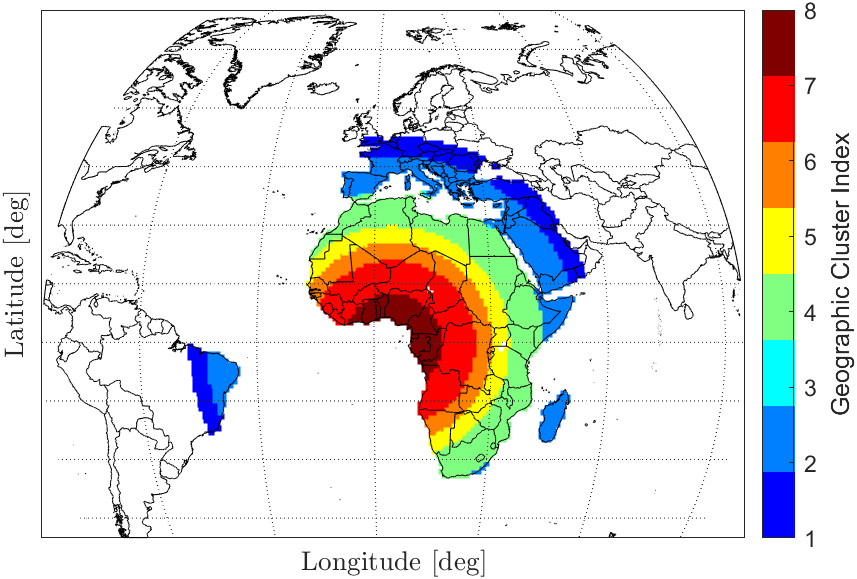}
  \label{fig:gc} }
  \caption{(a) SAT antenna aperture gain 3 dB footprint; (b) elevation angle of SAT as seen from Earth; (c) geographic clusters obtained with selected $G_s$ and $\psi_s$ resolution.}
\end{figure}

\subsection{Aggregation over the \gls{SATFP}}\label{Sec:AgglargeArea}
In case the aggregation area is larger than a single city, different locations on Earth's surface see the \gls{SAT} under different elevation angles. This causes the interference contribution from the same \gls{BS} model to be different based on latitude and longitude location \cite{mmWaveModellling0}. To obtain the total aggregated interference, we first identify the regions that share similar link budget parameters towards the \gls{SAT}, i.e., the same elevation angle and the same \gls{SAT} antenna gain. These regions are called Geographic Clusters (GC).

Let us consider the example of a \gls{SAT} occupying an orbital location (0N, 5E) in a geostationary orbit and having the uplink antenna pointing at Nadir. Also, assume that the locations on Earth's surface are discretized by defining a tessellation of pixels of $1$ deg along latitude and longitude. Fig~.\ref{fig:satgain} shows the pixels inside the 3 dB \gls{SATFP} assuming the antenna pattern model as defined in \cite{iturs672}, with \gls{SAT} gain $G_s$ quantized in steps of 1 dB. \textcolor{black}{The maximum gain and 3 dB beamwidth (22 dBi and 15$^\circ$ respectively) are taken from ITU WP4 discussions and, assuming a parabolic antenna mounted on the satellite \cite{orfanidis}}.
Assume that the set of GCs are $\mathcal{C} = \{{C}_{1}, {C}_{2}, \cdots, C_\upsilon,\cdots,{C}_{\Upsilon} \}$, where ${C}_{\upsilon}$ denotes the ${\upsilon}$-th CG. The number of GC is $|\mathcal{C}|=\Upsilon$. Now, the GCs inside \gls{SATFP} will observe the \gls{SAT} under different angles identified by azimuth and elevation pairs $\boldsymbol{\vartheta}_s^{\upsilon} = (\phi_s^{\upsilon}, \psi_s^{\upsilon})$, as depicted in Fig~.\ref{fig:satelev}.  Since the effect of the azimuth, $\phi_s^{\upsilon}$ is averaged out due to the fact that the \gls{BS}s are assumed to be randomly oriented throughout a large region (see Sec.~\ref{Sec:Array}), only the impact of the \gls{SAT} elevation is of interest. In Fig~.\ref{fig:satelev}, the elevation angle of the \gls{SAT} seen from Earth is depicted for each pixel with an angle quantization of $10$ deg. 
Each GC can thus be represented as a unique pair $(G_s^{\upsilon}, \psi_s^{\upsilon})$. The GCs obtained with the selected resolution are reported in Fig~.\ref{fig:gc}.  
To compute the aggregated interference coming from the \gls{SATFP}, for each GC we find the CF of its interference contribution following the procedure detailed in Sec.~\ref{Sec:InterfAnalysis}. 
Then we aggregate the CFs of all the GCs to find the overall result as
\begin{equation}\label{eq:footprint}
    \Phi_{FP}(\omega) = \prod_{\upsilon=1}^{\Upsilon}  \Phi_{\upsilon}(\omega)
\end{equation}
where $\Phi_{\upsilon}(\omega)$ is the CF of interference from the $\upsilon$-th GC, computed with \eqref{eq:agg_city}, using the corresponding \gls{SAT} Rx gain on each GC $G_s^{\upsilon}$ and the average number of \gls{BS}s in the $\upsilon$-th GC $Q_{\upsilon} = \lambda \rho  F_{T} R_a R_b S_{\upsilon}$. The latter is the product of the GC area $S_{\upsilon}$, the average density of the macro \gls{BS}s $\lambda$, the \gls{BS} loading factor $\rho$, \textcolor{black}{\gls{TDD} activity factor $F_{T}$}, while $R_a$ and $R_b$ are parameters defined by ITU \cite{annex4.4}, to establish a bond between amounts of BS and large-scale land areas in the order of SATFP, where $R_b$ is the percentage of built area, and $R_a$ is the percentage of the area from a certain type, e.g., urban, suburban, and etc. Here we consider urban area type. 
\begin{figure*}[t!]
    \hspace{-1cm}
    \hspace{0.25cm}
    \subfloat{\includegraphics[scale=0.45]{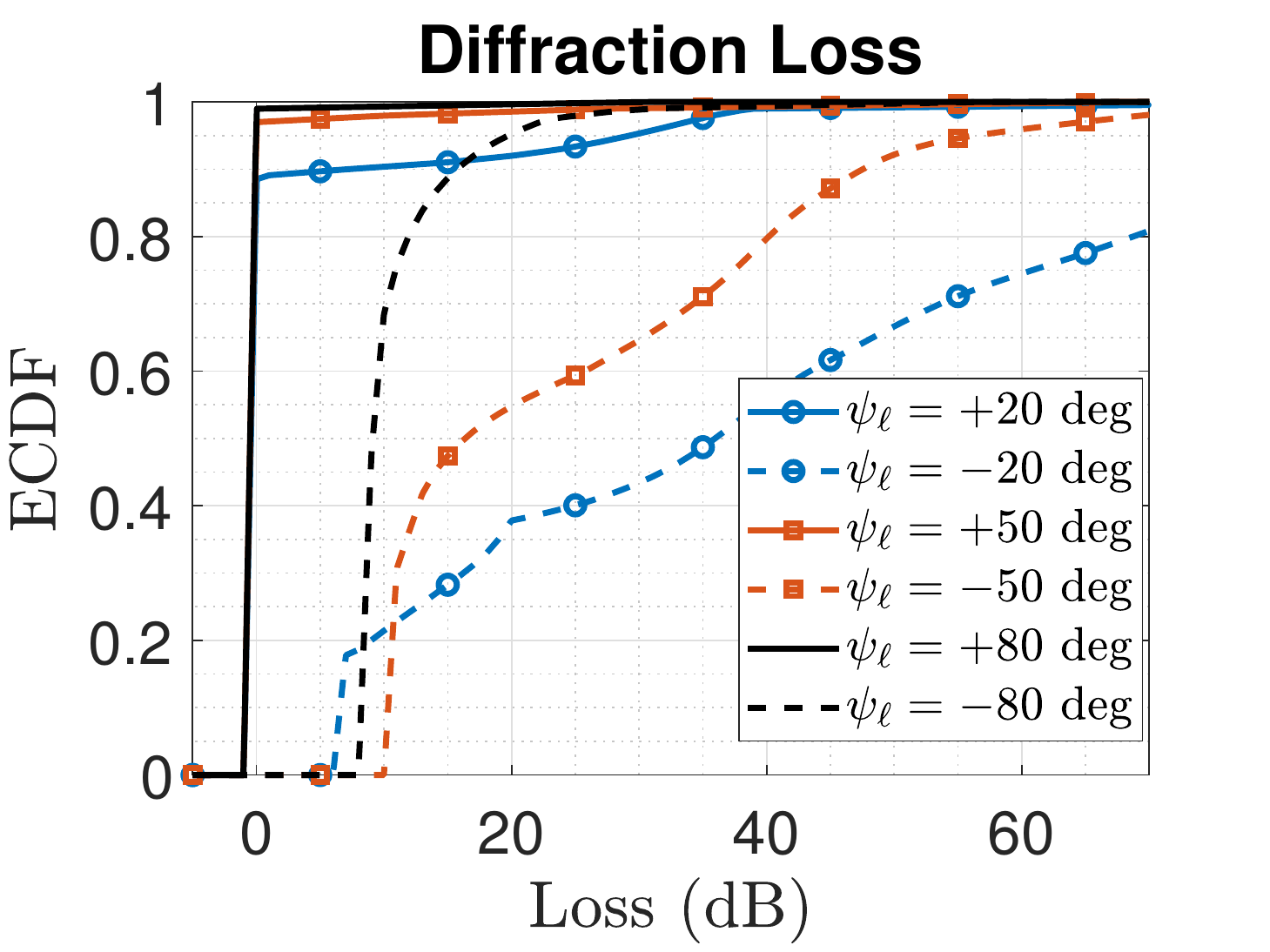}\label{fig:Diff}}
    \hspace{-0.5cm}
    \subfloat{\includegraphics[scale=0.45]{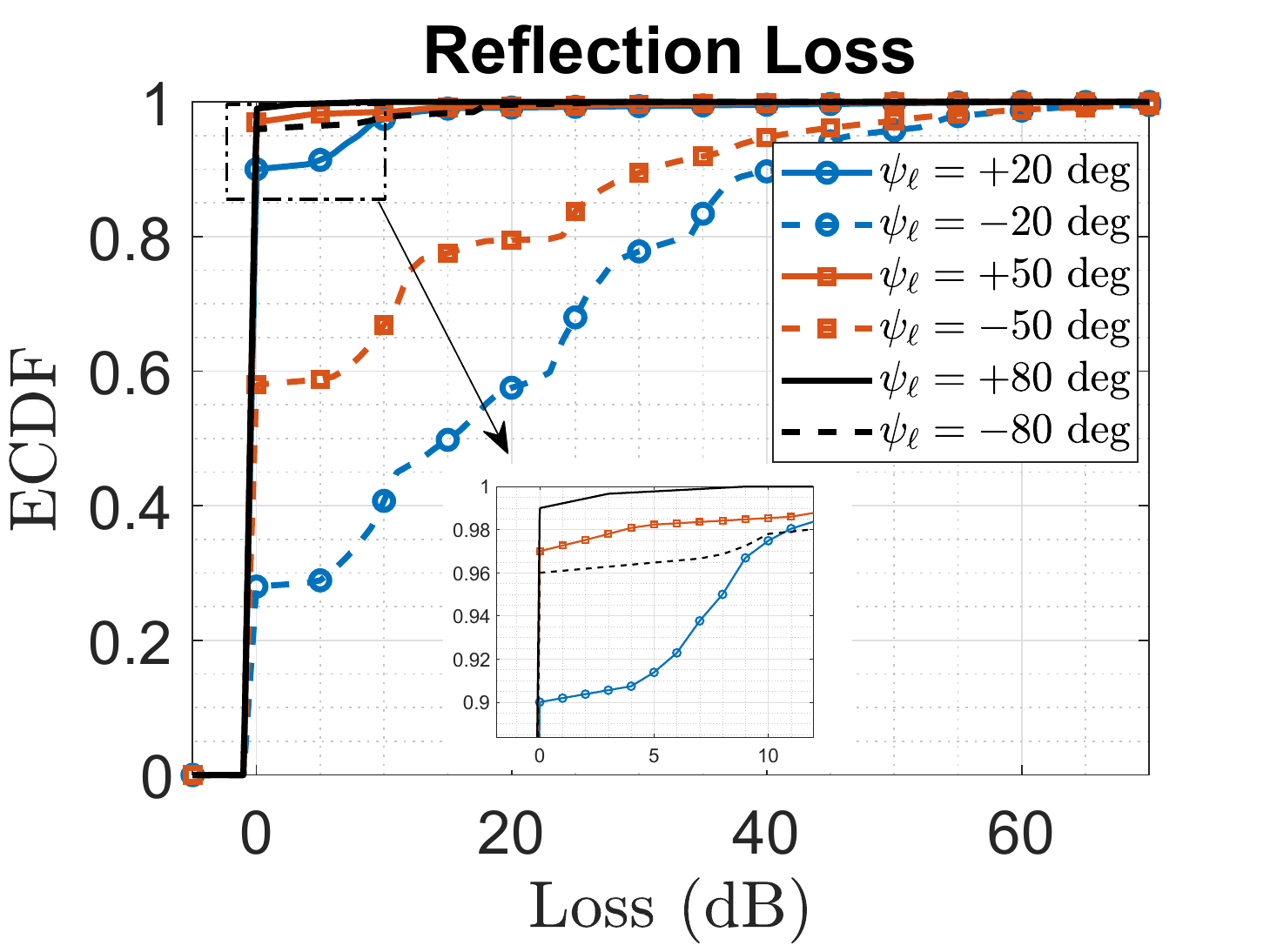}\label{fig:Refl}}
    \hspace{-0.5cm}
    \subfloat{\includegraphics[scale=0.45]{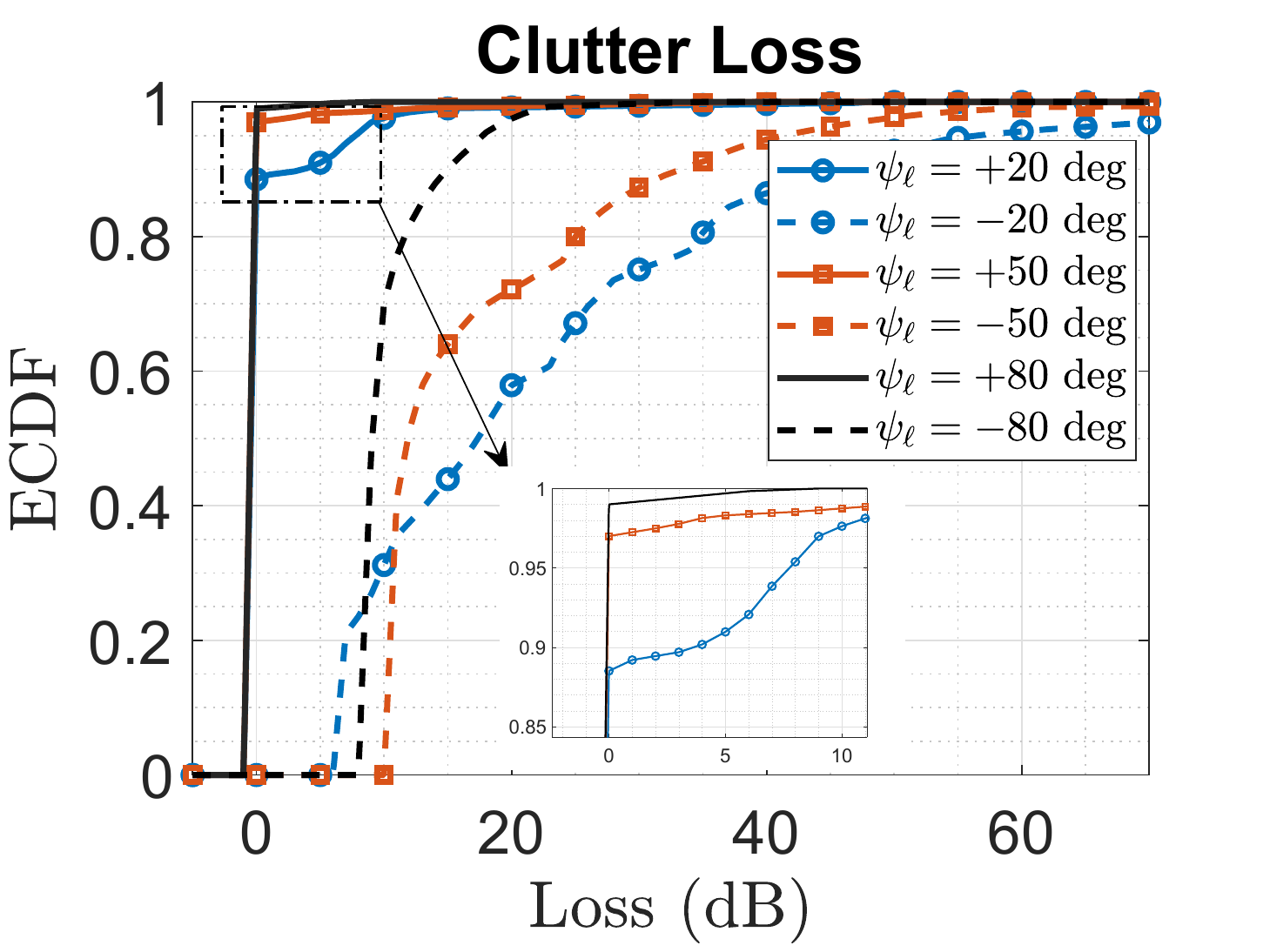}\label{fig:Clut}}
    \caption{Empirical CDF (ECDF) of diffraction loss, reflection loss and clutter loss for the \gls{SAT} elevation angles $\psi_s \in\{20,50,80 \}$ deg for positive and negative modes, with \gls{BS} height $h_{BS} = 6m$, for the city of Milan using the method in \cite{RIVA}.\label{fig:CL}}
\end{figure*}
\subsection{Interference protection criterion}
Given the interference power $I$ and signal bandwidth $B$, the interference protection criterion, is based on the \gls{INR} defined as \cite{iturs1323}:
\begin{equation}
    \textrm{INR} = \frac{I}{B N_0},
\end{equation}
where $N_0 = K_B T_\textrm{sys}$ is the noise \gls{PSD} with $K_B$ as the Boltzmann constant and $T_\textrm{sys}$ is the SAT Rx system temperature \cite{SATCOMM}. The SAT is protected from interference whenever interference threshold criterion $\textrm{INR}\leq \textrm{INR}_{th}$, with $\textrm{INR}_{th}$ being a threshold specified by satellite regulators. In \cite{WRC19}, the $T_\textrm{sys}$ is given for \gls{mmW} in the range of $400-950 $ K depending on different parameters, while the calculations in \cite{Guarnieri}, demonstrate system thermal noise of around $800$ K in the C-band.
%

\section{Stochastic Array Gain}\label{Sec:Array}
This section details the modeling of the array gain $G_a\left( \boldsymbol{\vartheta}  \lvert \boldsymbol{\vartheta}_k, \boldsymbol{\Theta}\right)$ and its PDF. We consider generic rectangular panels for each BS sector, each configured with $N_V$ vertical and $N_H$ horizontal antennas. The \gls{EIRP} is therefore defined as 
\begin{align}
    \textrm{EIRP}= \frac{P_T N_V^2  N_H^2}{\eta}, 
\end{align}
where $\eta$ is the sub-array size (number of antennas connected to a single RF chain) in a hybrid digital-analog antenna array, and the Tx power $P_T$, is related to a single RF chain (i.e., a single \gls{PA}) \cite{iturm2101,iturm2292}, while $\eta=1$ corresponds to a fully digital antenna array. A feeder loss can be added to further reduce the effective EIRP as indicated in \cite{itut_k_supp16}. The horizontal and vertical gains assigned toward a generic azimuth $\phi$ and elevation $\psi$ steering toward the $k$-th UE are 
\begin{align}
    &G_{H}(\phi|\phi_k) = \left|\mathbf{b}^H(\phi_k - \varrho) \mathbf{a}(\phi - \varrho) \right|^2{D_{H}\left(\phi-\varrho\right)},\\
    & G_{V}( \psi|\psi_k) = \left|\mathbf{b}^H(\psi_k) \mathbf{a}( \psi+\psi_{tilt}) \right|^2{D_{V}\left( \psi+\psi_{\text {tilt }}\right)},
\end{align}
where: \textit{(i)} $D_{H}(\phi)$ and $D_{V}(\psi)$ are the horizontal and vertical element directivity gains \cite{3GPP}, respectively, \textit{(ii)} $\mathbf{a}(\phi)\in \mathbb{C}^{N_H \times 1}$ and $\mathbf{a}(\psi)\in \mathbb{C}^{N_V \times 1}$ are the horizontal and vertical ULA response vectors, respectively, \textit{(iii)} $\mathbf{b}(\phi)\in \mathbb{C}^{N_H \times 1}$ and $\mathbf{b}(\psi)\in \mathbb{C}^{N_V \times 1}$ are the conventional horizontal and vertical beamforming, $\mathbf{b}^{\mathrm{H}}$ indicates the hermitian of vector $\textbf{b}$, and $\varrho$ is the orientation of the serving \gls{BS} panel that is perceived by the \gls{SAT} as $\varrho\sim U[0,2 \pi]$. Note that $\varrho$ corresponds to any \gls{BS} panel that observes the target azimuth $\phi$ within the \gls{EM} shielding limit while serving the $k$-th UE, as $|\phi_k - \varrho|\leq \phi_{sh}$, where $\phi_{sh}=60$ deg. With such an assumption, it is apparent that only one of the panels of the macro \gls{BS} is capable of interfering with $\phi$ while serving a UE at $\phi_k$. In the case of micro \gls{BS}, depending on the number of the \gls{BS}s and their orientations, one or multiple \gls{BS}s might be interfering. The total gain is
\begin{equation}
    G_a\left( \boldsymbol{\vartheta}  \lvert \boldsymbol{\vartheta}_k,\boldsymbol{\Theta}\right)=G_{V}\left( \psi  \lvert \psi_{k}\right) G_{H}\left(\phi  \lvert \phi_{k}\right).
\end{equation}
The corresponding array gain for $\ell$-th \textcolor{black}{interference mode} with AoD $\boldsymbol{\vartheta}_{\ell}$ is denoted as $G_a(\boldsymbol{\vartheta}_{\ell}|\boldsymbol{\vartheta}_{k},\boldsymbol{\Theta})$, that for the direct \gls{BS}-SAT link would be $G_a(\boldsymbol{\vartheta}_{s}|\boldsymbol{\vartheta}_{k},\boldsymbol{\Theta})$.
The \gls{PDF} of the array gain $\mathcal{P}(G_a|\boldsymbol{\vartheta}_{\ell},\boldsymbol{\Theta})$ is achieved by Monte-Carlo simulations, given the random $\boldsymbol{\vartheta}_k$. This \gls{PDF} is for the array gain toward a single UE, while the number of the served UEs affects the Tx power and the \gls{BS} loading factor, rather than the array gain. Note that the random AoD of the k-th UE $\boldsymbol{\vartheta}_k$, is inherently a function of environment parameters $\boldsymbol{\Theta}$, as it depends on the cell size, random 2D position of the UEs within the cell, the distribution of the UEs height, and the distribution of the \gls{BS} height. 
\section{Stochastic Clutter Loss}\label{Sec:CL}
Clutter is the term herein used to indicate objects that are on the Earth's surface, but are not the terrain itself, i.e. buildings and vegetation. Clutter loss $A_c(\boldsymbol{\vartheta}_{\ell} \lvert \boldsymbol{\Theta})$ consists of reflection loss $A_r(\boldsymbol{\vartheta}_{\ell} \lvert \boldsymbol{\Theta})$ and diffraction loss $A_d(\boldsymbol{\vartheta}_{\ell} \lvert \boldsymbol{\Theta})$, properly combined as indicated in \cite{iturp2402}.   
Quantifying clutter loss is not trivial, since it is strongly dependent on both environment and geometry of the link of interest. The target \gls{AoD} also plays a crucial role \cite{CL_LowElev,CL_Elev}. One possible approach is to use a ray tracer to evaluate clutter loss via deterministic simulation, but the computational load makes this method applicable to limited areas only, yielding site-specific results that are not general enough \cite{CL_LowElev}. As previously mentioned, ITU recommendation \cite{iturp2402} contains guidelines to perform stochastic Monte Carlo simulation of clutter loss statistics, making use of environment geometrical data and stochastic geometry to evaluate the \gls{CDF} of clutter loss at a given elevation angle $\psi_s$. Some of the stochastic parameters of the environment that serve as the input for this method are BS height, the material of the buildings/ground, and some specific percentiles of the inter-building distances and buildings' height.

Although the statistical nature of the approach in \cite{iturp2402} is very much suited to our model, there are some limitations that can affect the applicability of this model: the model is not considered valid below 10 GHz, a limited number of reflections and diffraction are considered (up to two), the reflection coefficients are not dependent on angles of incidence and polarization. Furthermore, the main drawback is that this method is designed for only the direct \gls{BS}-SAT link, and it does not provide distinct clutter loss statistics for different \textcolor{black}{interference modes}. This is while the SMI method require a distinct clutter loss $A_c(\boldsymbol{\vartheta}_{\ell}\lvert\boldsymbol{\Theta})$ corresponding to the $\ell$-th propagation mode. In this regard, one could resort to the model proposed in \cite{RIVA} as an extension of \cite{iturp2402}, where the aforementioned limitations are overcome, considering \textit{positive} and \textit{negative} interference modes. Positive modes are all the modes that leave the \gls{BS} upward with elevation $+\psi_s$ and negative modes are all the modes that leave the \gls{BS} downward with elevation $-\psi_s$, i.e. ground reflection modes. Fig.~\ref{fig:CL} shows the clutter-loss, diffraction-loss and reflection-loss, given the extracted geometrical statistics of Milan, when the \gls{BS} is located at $h_{BS} = 6$m  height.

\textit{Remark:} Being more specific regarding diffraction and reflection, it can be understood that the \gls{GSMI} method, in fact, mimics the effect of diffraction loss, in a hard decision manner, i.e., a ray is in \gls{LOS} mode or fully blocked in a stochastic manner. However, in some cases, reflection loss is dominant as seen in Fig \ref{fig:CL}. Thus, in order to take the reflection loss into account (if it is available), one can repeat the same procedures of GSMI method, by replacing $G_c(\boldsymbol{\vartheta}_{\ell} \lvert \boldsymbol{\Theta})$ and $\mathcal{P}(G_c  \lvert \boldsymbol{\vartheta}_{\ell}, \boldsymbol{\Theta})$ with reflection gain $G_r(\boldsymbol{\vartheta}_{\ell} \lvert \boldsymbol{\Theta}) = A_r(\boldsymbol{\vartheta}_{\ell} \lvert \boldsymbol{\Theta})^{-1}$ and its PDF $\mathcal{P}(G_r  \lvert \boldsymbol{\vartheta}_{\ell}, \boldsymbol{\Theta})$ in relations \eqref{eq:interf_power} and \eqref{eq:PDF_I}, respectively.

\begin{figure}[!t]
    \centering
    \subfloat[][]{\includegraphics[width=0.85\columnwidth]{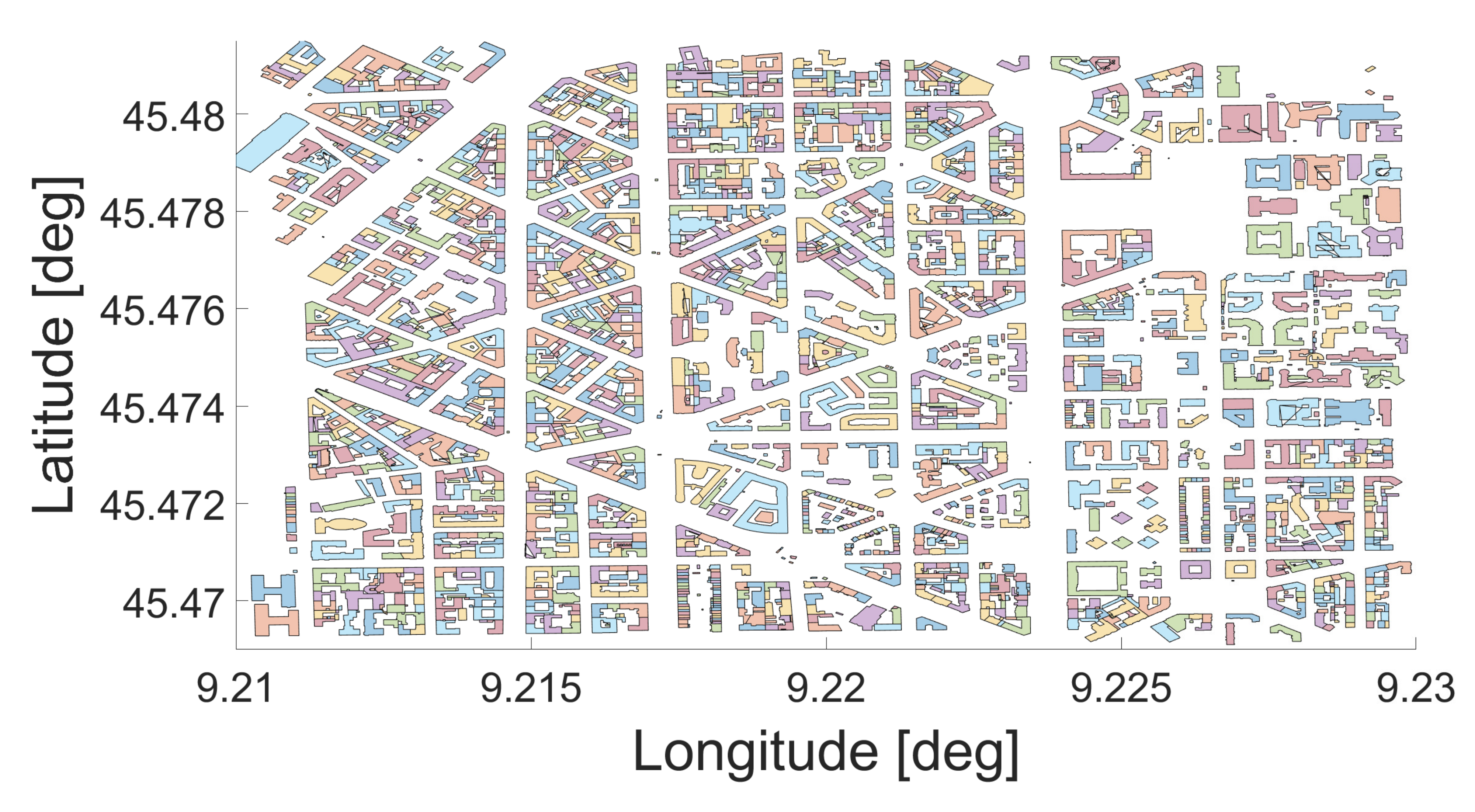}\label{fig:original_data}}\\\vspace{-0.3cm}
    \subfloat[][]{\includegraphics[width=0.85\columnwidth]{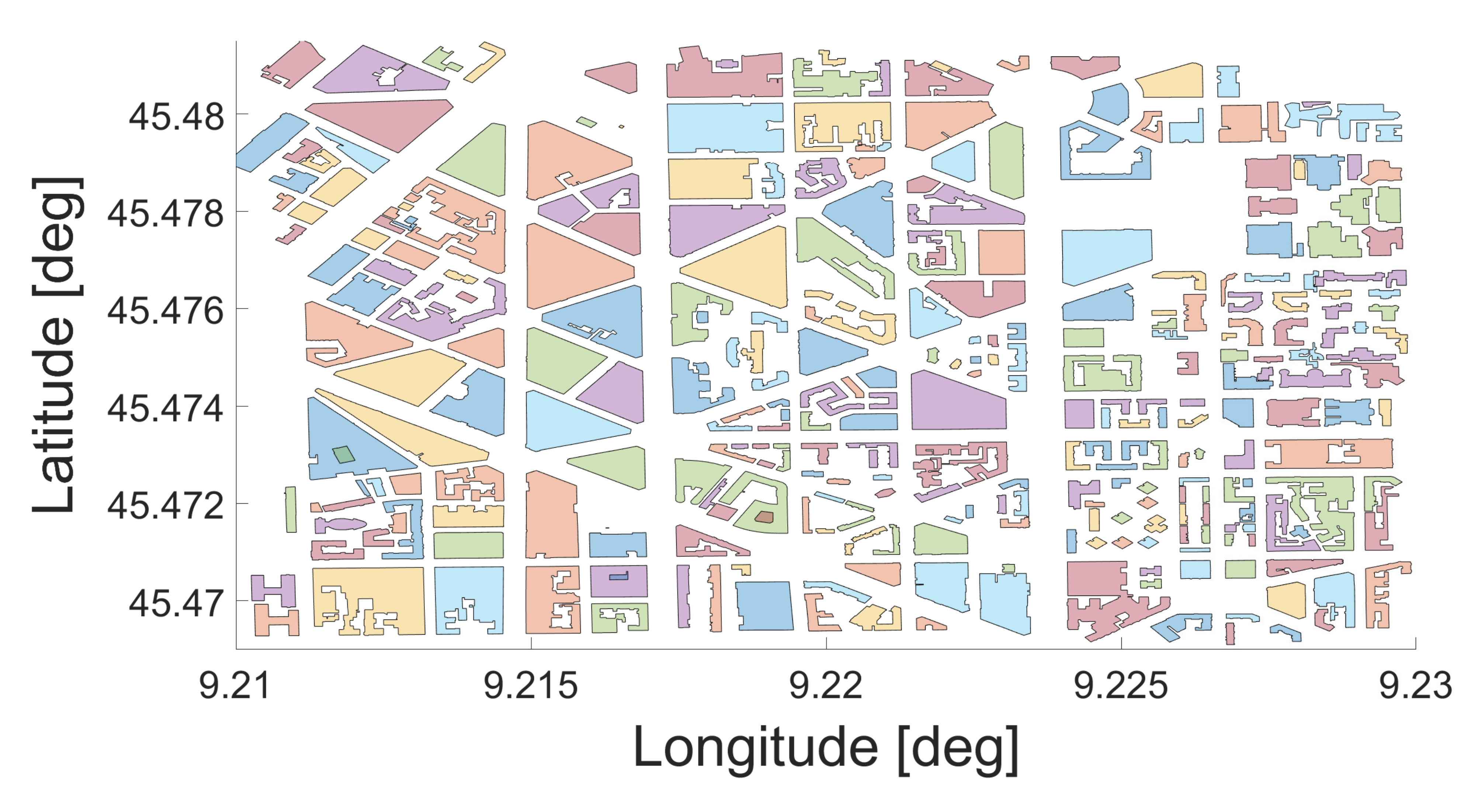}\label{fig:processed_data}}\\\vspace{-0.3cm}
    \subfloat[][]{\includegraphics[width=0.85\columnwidth]{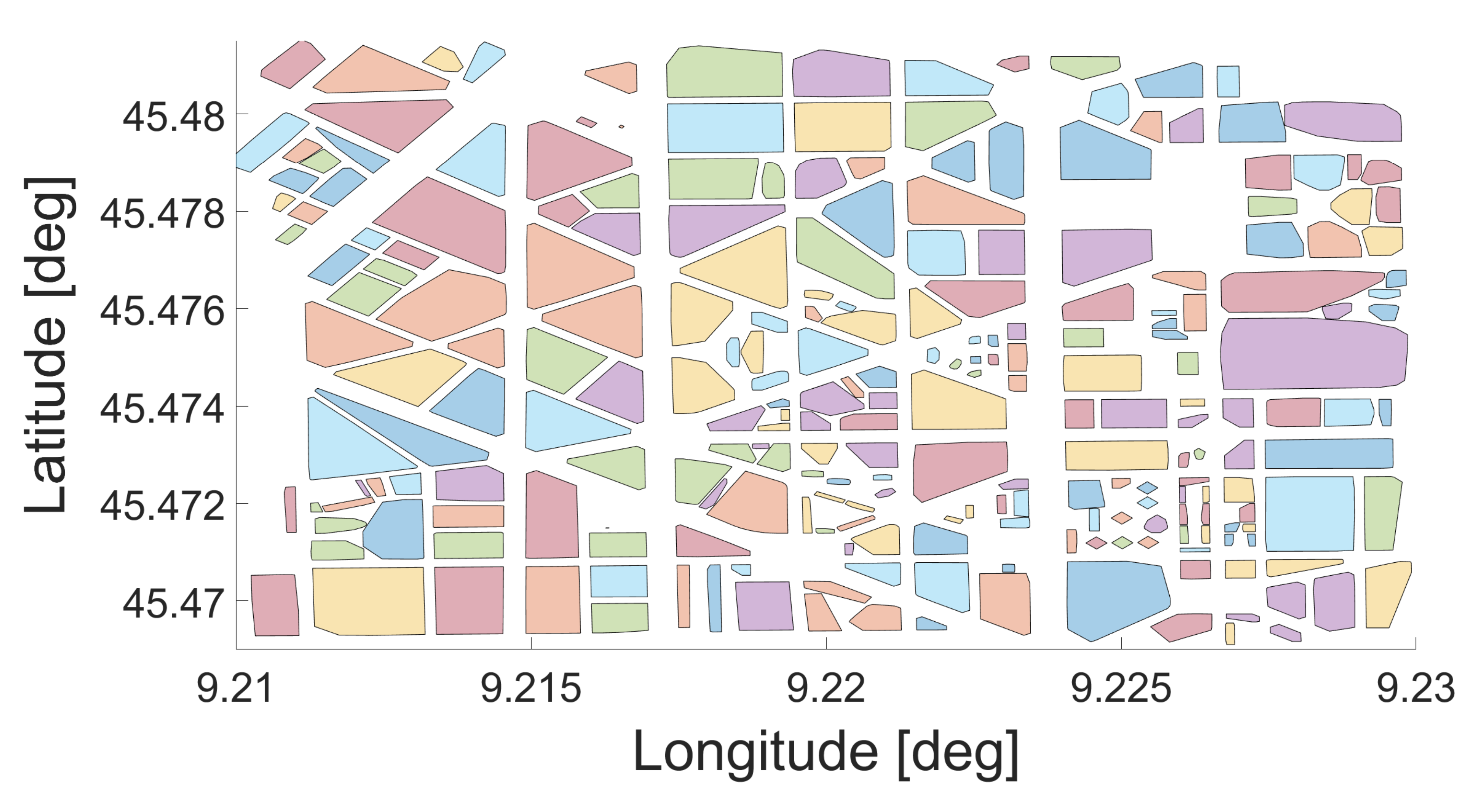}\label{fig:convex_data}}
    \caption{Dataset processing chain: (a) original dataset (city of Milan), (b) merging process, (c) convex shape approximation.\label{fig:building_process}}
\end{figure}
\section{Geometrical Statistics}\label{Sec:geom_stat}

\begin{figure*}[t!]
    \centering
    \hspace{-1cm}
    \subfloat{\includegraphics[scale=0.42]{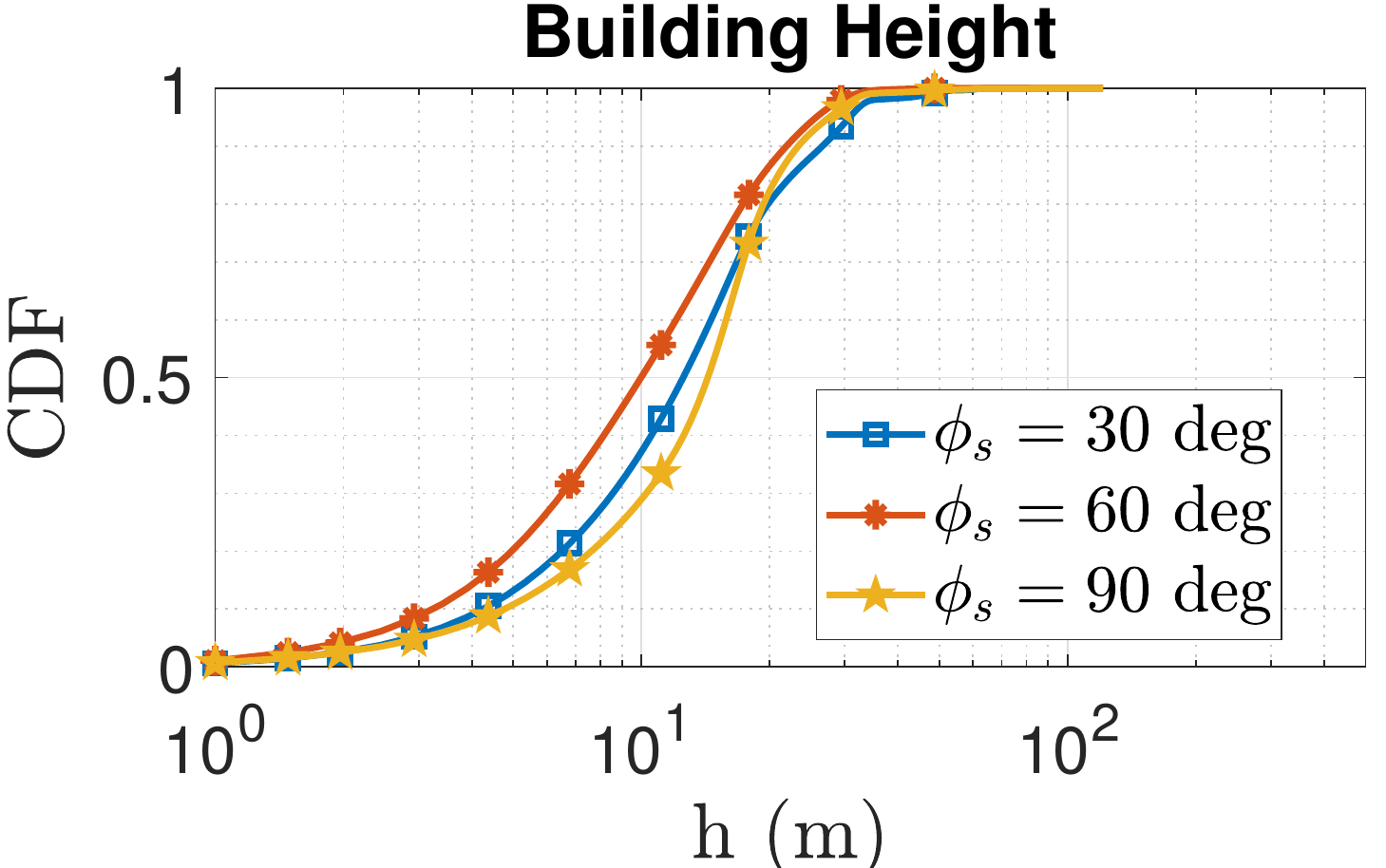}\label{fig:Height}}
    \hspace{+0cm}
    \subfloat{\includegraphics[scale=0.42]{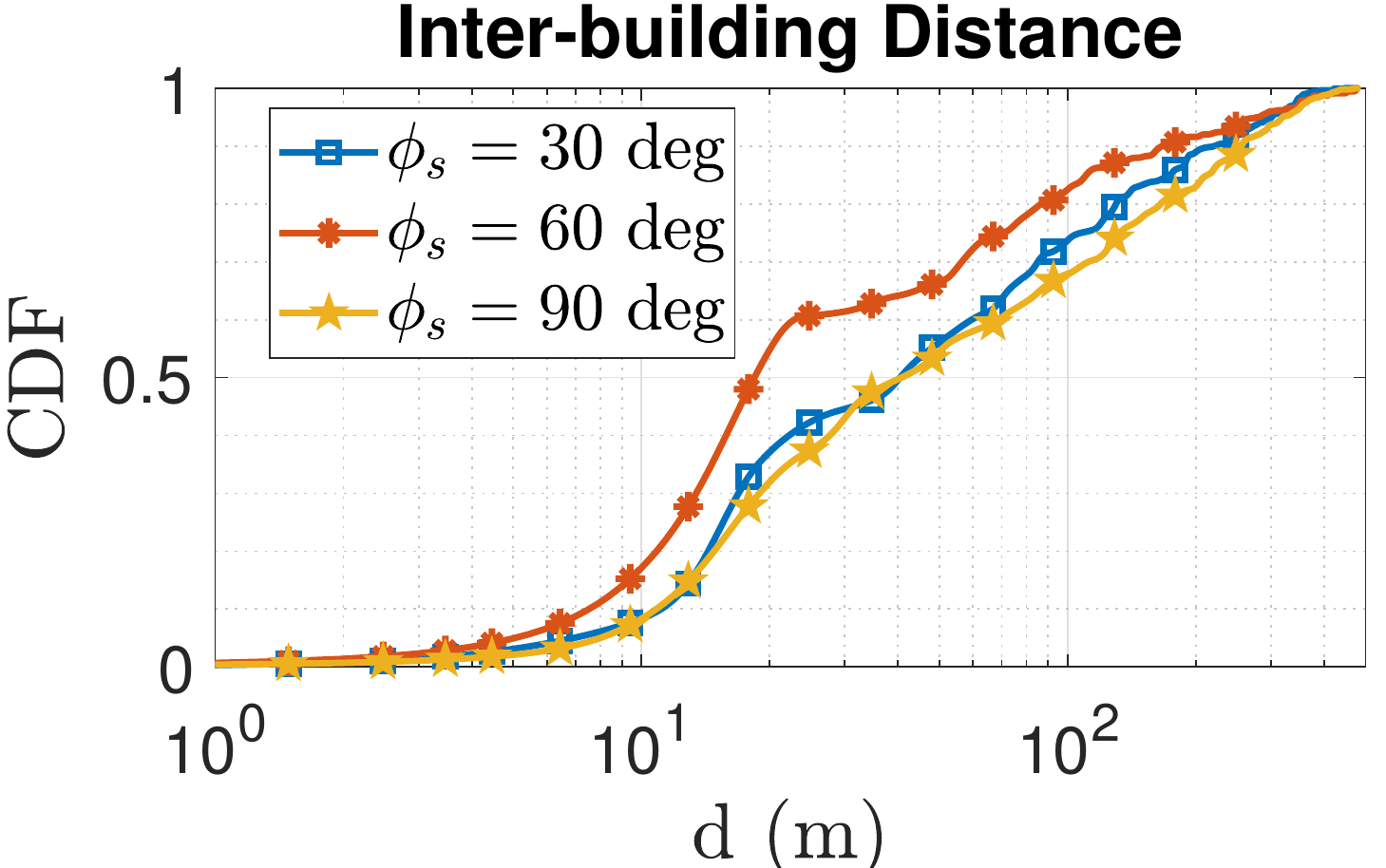}\label{fig:Distance}}
    \hspace{+0cm}
    \subfloat{\includegraphics[scale=0.42]{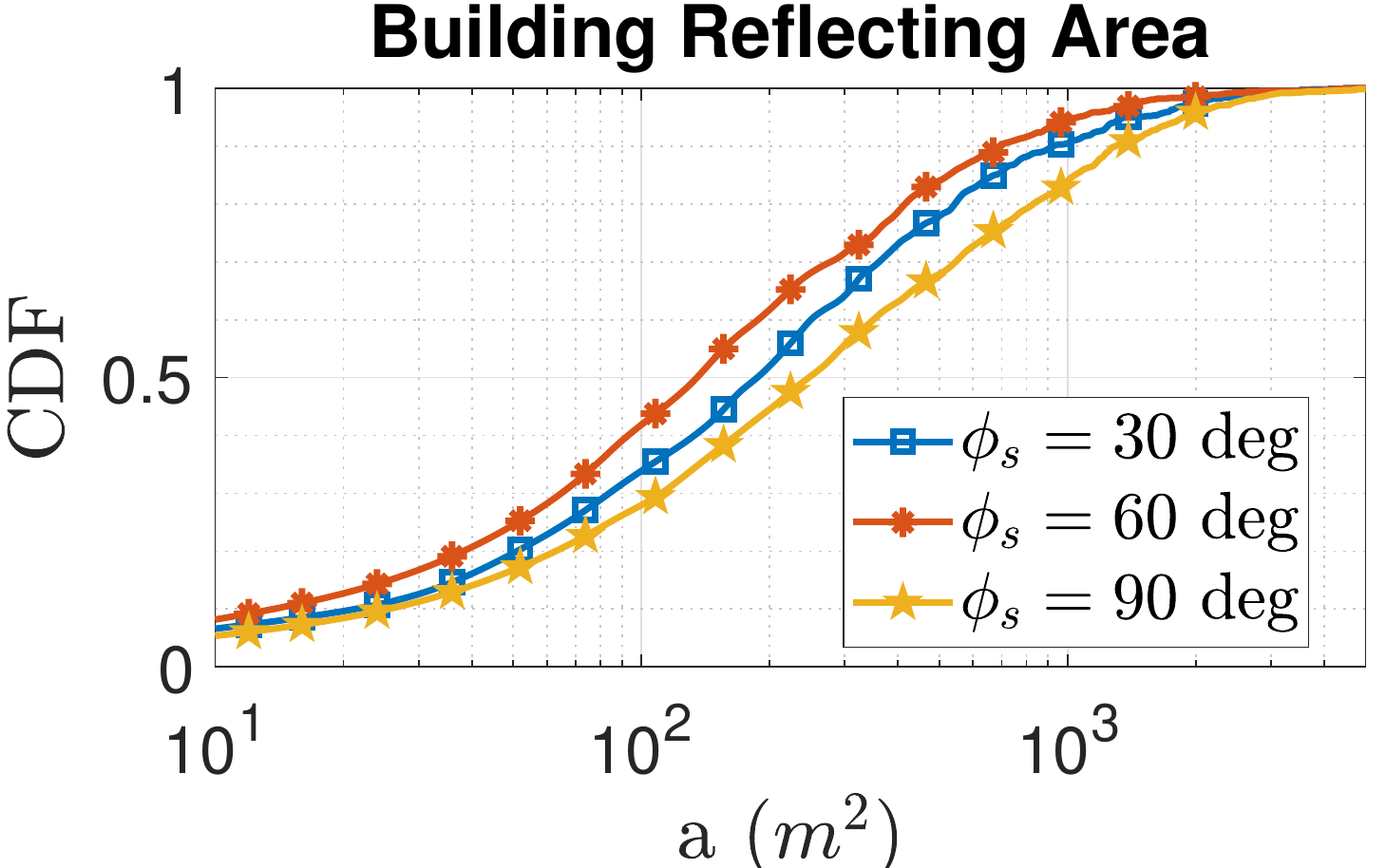}\label{fig:Area}}
    \caption{\gls{CDF} of the geometrical parameters, extracted for city of Milan, given $\phi_s = \{  30,60,90\}$ deg\label{fig:GeomDist}.}
\end{figure*}
As shown in Section~\ref{Sec:SysModel}, the set of geometrical parameters $\boldsymbol{\Theta}$ is required to calculate the interference power. These parameters affect both the array gain and clutter loss (see Sec. \ref{Sec:Array} and Sec.~\ref{Sec:CL}). Furthermore, they are necessary to compute the occurrence probability $\mathcal{P}_{\ell}(\boldsymbol{\Theta})$ in the GSMI method as \eqref{eq:deconditionOcurrence}.
These geometrical statistics used are extracted using a pseudo-3D or 2.5D approach \cite{25D_1,25D_2}, where the 3D geometry is split into 2D cross-sections along the SAT azimuth $\phi_s$. The city of Milan is taken as a reference, and we generate \textit{(i)} the \gls{PDF}s of the buildings' heights, \textit{(ii)} the \gls{PDF}s of the reflection area of each building's facade, and \textit{(iii)} the \gls{PDF}s of buildings' inter-distance (i.e, streets' widths), each on a regular azimuth grid with quantization step of $\Delta\phi = 5$ deg, using relevant public datasets with further processing \cite{geo_portale}. Figure~\ref{fig:original_data} shows an exemplary portion of Milan from the original dataset, where each polygon defines a specific detail of a building with a particular height. Such details are not required, as we are interested in representing only the external buildings' facades, and some merging processes can be applied to reduce the complexity of the environment while maintaining useful geometrical information. The convexification process is shown in Figure~\ref{fig:original_data} to \ref{fig:convex_data}, where the final convex  polygons are associated with average heights and widths, retrieved for each merged building. Although not reported here, it can be shown that the merging-plus-convex approximation of the buildings' geometry preserves the facades' area. 
\begin{figure}[t!]
    \centering
    \includegraphics[width=0.95\columnwidth]{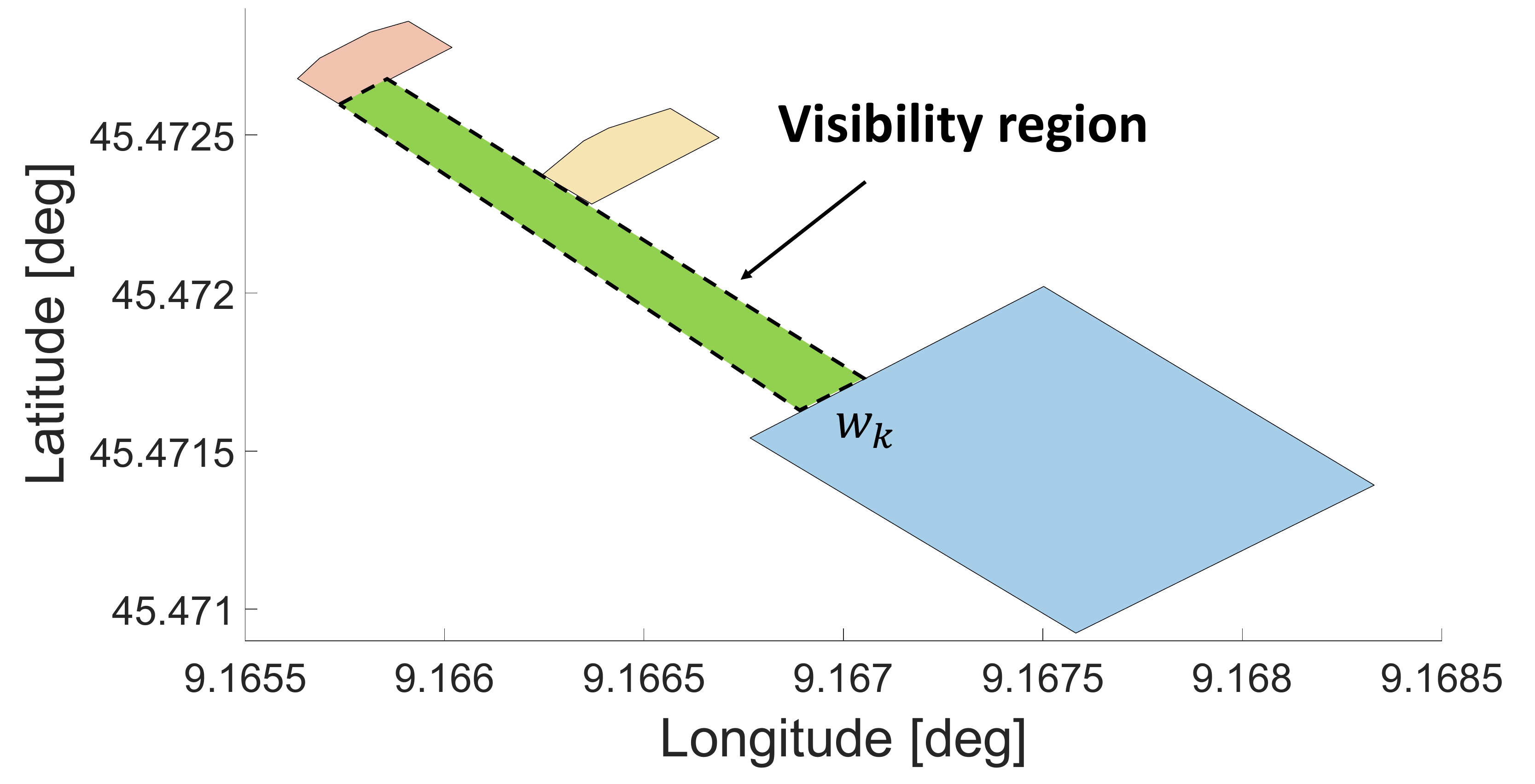}
    \caption{Inter-distance between the buildings having 2 parallel walls. The green area identifies the visibility region.}
    \label{fig:vis_region}
\end{figure}

From the simplified dataset, we generate the PDFs for each SAT azimuth angle $\phi_s$. The buildings' heights are discretized at $\Delta h=5$ m and the PDF $\mathcal{P}(h|\phi_s)$ is derived from the histogram.
%
%
Each entry of the histogram is a weighted sum of all the building's facades of height $h_i$ perpendicular to the azimuth satellite direction $\phi_s$, (i.e., that can effectively contribute to the clutter loss)
The PDF of the buildings' reflecting area $\mathcal{P}(a|\phi_s)$ is evaluated similarly, by using the histogram of occurrence of a certain reflection area $a$, discretized with a step of $\Delta a = 20$ m$^2$. Differently, the PDF of the buildings' inter-distance is weighted by the effective \textit{visibility region} between adjacent buildings, as illustrated in Fig~.\ref{fig:vis_region}.
The PDF is again approximated as
\begin{equation}\label{eq:distance_prob}
    \mathcal{P}(d|\phi_s) \approx \frac{\mathbf{d}(\phi_s)}{\sum_{j=1}^{N_h}[\mathbf{d}(\phi_s)]_j},
\end{equation}
where $\mathbf{d}(\phi_s)$ is the inter-distance histogram quantized with step $\Delta d=5$ m, but the $j$-th histogram element $[\mathbf{d}(\phi_s)]_j$ is now computed as
\begin{equation}
    [\mathbf{d}(\phi_s)]_j = \sum_{k\in\mathcal{W}_j(\phi_s)} w_k h_k,
\end{equation}
where $w_k$ is the width of the visibility region of two adjacent buildings with parallel facades and $h_k$ is the average height of the two involved facades. The set $\mathcal{W}_j(\phi_s)$, therefore, spans all the pairs of parallel facades at distance $d_j$ and perpendicular to $\phi_s$. Fig.~\ref{fig:GeomDist} shows the distribution of building's height $h$, inter-building distance $d$, and buildings' reflection area $a$, given some exemplary \gls{SAT} azimuth $\phi_s = \{30,60,90 \}$ deg.

\section{Numerical Results}\label{Sec:NumericalResults}
This section shows the numerical aggregated interference power density for the city of Milan and the \gls{SATFP}.

\subsection{Simulation setup}

The \gls{BS} arrangement is the same depicted in in Fig~.\ref{fig:Cell_Shapes} (Section \ref{Sec:SysModel}). For each macro cell, we consider 3 single-sector micro BS, for a total of 9 micro BS for each macro BS. Micro BSs are placed at $h_{BS}=6$ m \cite{iturm2292,mmWaveModellling7}, while macro BSs at $h_{BS}=\textrm{max}(h,6)$ m, where $h$ is the height of the tallest building in the 3 macro cells pertaining to the same macro BS. The macro cell radius considered is $d_{c} = 300$ m \cite{iturm2292} and the macro BS density is therefore $\lambda = 1/(3S_c)$, where $S_c$ is the macro cell's area. The micro cell radius is assumed to be $d_{m} = d_{c}/4$. Each micro \gls{BS} serves UEs within $d_{m}$, while the rest are served by the macro BS.

\begin{figure}[tb!]
\hspace{-1cm}
    \centering    \includegraphics[width=0.9\columnwidth, ]{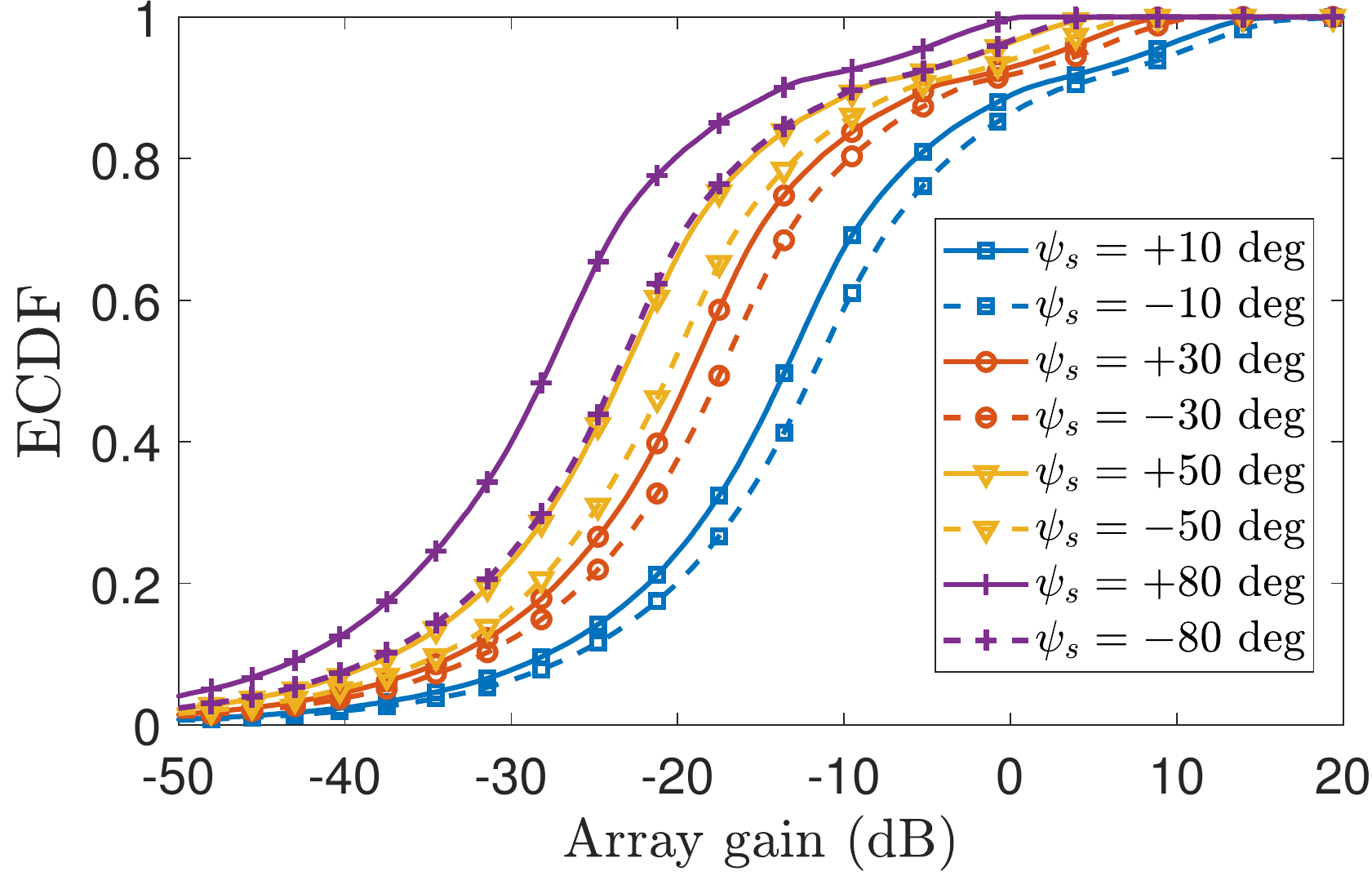}
    \caption{\gls{CDF} of the Array gain for a macro \gls{BS}, serving UEs inside the buildings, distributed vertically uniform over the height of the building, using the array configuration 1, in Table \ref{tab:ArrayConf}. }
    \label{fig:Array_Gain_CDF}
\end{figure}

The \gls{UE}s are considered to be randomly located either on the ground (outdoor UEs) and inside the buildings (indoor UEs), according to a 2D random distribution with spatial density $\lambda_{UE}$ [UE/m$^2$] (on the ground plane). Outdoor UEs are assumed to have a constant height $h_{UE}=1.5$ meter. Indoor UEs' height is assumed to be $h_{UE} \sim U[1.5,h]$ meters, where $h$ follows the distribution of the buildings' height.
\begin{table}[t!]
\centering
\caption{Array configurations}
\label{tab:ArrayConf}
\renewcommand{\arraystretch}{1.2}
\begin{tabular}{c|cc|cc|}
\cline{2-5}
& \multicolumn{2}{c|}{Config. 1}     & \multicolumn{2}{c|}{Config. 2}     \\ \cline{2-5} 
& \multicolumn{1}{c|}{macro} & micro & \multicolumn{1}{c|}{macro} & micro \\ \hline
\multicolumn{1}{|c|}{$N_H$}          & \multicolumn{1}{c|}{8}    & 4     & \multicolumn{1}{c|}{8}    & 8     \\ \hline
\multicolumn{1}{|c|}{$N_V$}          & \multicolumn{1}{c|}{8}    & 8    & \multicolumn{1}{c|}{16}    & 8    \\ \hline
\multicolumn{1}{|c|}{$\eta$}         & \multicolumn{1}{c|}{1}     & 1     & \multicolumn{1}{c|}{2}     & 2     \\ \hline
\multicolumn{1}{|c|}{$A_f$ (dB)}         & \multicolumn{1}{c|}{3}     & 3     & \multicolumn{1}{c|}{3}     & 3     \\ \hline
\multicolumn{1}{|c|}{$P_T$  (dBm)} & \multicolumn{1}{c|}{25}    & 19    & \multicolumn{1}{c|}{22}    & 16    \\ \hline
\multicolumn{1}{|c|}{\textrm{EIRP} (dBm)}     & \multicolumn{1}{c|}{58}    & 46    & \multicolumn{1}{c|}{58}    & 46    \\ \hline
\multicolumn{1}{|c|}{$\psi_{3\textrm{dB}}$ (deg) \cite{3GPP}}         & \multicolumn{1}{c|}{65}     & 65     & \multicolumn{1}{c|}{65}     & 65     \\ \hline
\multicolumn{1}{|c|}{$\phi_{3\textrm{dB}}$ (deg) \cite{3GPP}}         & \multicolumn{1}{c|}{65}     & 65     & \multicolumn{1}{c|}{65}     & 65     \\ \hline
\multicolumn{1}{|c|}{$\psi_{\textrm{tilt}}$ (deg) }         & \multicolumn{1}{c|}{-10}     & -10     & \multicolumn{1}{c|}{-10}     & -10     \\ \hline
\multicolumn{1}{|c|}{$G_e$ (dBi) \cite{3GPP}}         & \multicolumn{1}{c|}{8}     & 8     & \multicolumn{1}{c|}{8}     & 8     \\ \hline
\end{tabular}
\end{table}
Table \ref{tab:ArrayConf} shows the array configurations used for macro and micro \gls{BS}s in this paper. The antenna element directivity model is based on \cite{iturm2101,3GPP}, with maximum element directivity gain $G_e$, vertical and horizontal fields of view of $\psi_{3\textrm{dB}}$ and $\phi_{3\textrm{dB}}$, respectively, and feeder loss $A_f$ (see Table \ref{tab:ArrayConf}).
Fig~.\ref{fig:Array_Gain_CDF}, shows an example of the \gls{CDF} of the array gain for the macro BS toward $\vartheta_{\ell}$, averaged over the distribution of the height of the buildings and the height of the UEs, when serving an indoor UE with array configuration 1 of table \ref{tab:ArrayConf}. It can be noticed that \textcolor{black}{interference modes} characterized by $\psi_{\ell} < 0$ (i.e., reflections from the ground) are characterized by a higher array gain, and, consequently, a larger interference contribution as the BS is tilted towards the ground.
Table.~\ref{tab:CalcParams} summarizes other relevant simulation parameters. 
\begin{table}[!tb]
\centering
\caption{Simulation parameters \label{tab:CalcParams}}
\renewcommand{\arraystretch}{1.2}
\begin{tabular}{|c|c|c|}
\hline
Parameter  Name                                                 & Parameter             & Value \\ \hline
SAT azimuth                                                       & $\phi_s$ (deg)        & 45    \\ \hline
central frequency                                                 & $F_c$ (GHz)           & 6     \\ \hline
SAT distance                                                      & $d_s$  (Km)          & 35000 \\ \hline
path loss                                                         & $A_{p}$ (dB)          & 199  \\ \hline
bandwidth                                                         & B  (MHz)              & 100   \\ \hline
Macro cell radius                                                         & $d_c$  (m)              & 300   \\ \hline
SAT Rx temperature \cite{Guarnieri}& $T_\textrm{sys}$ (Kelvin)  & 800   \\ \hline
threshold INR at $80\%$ \cite{annex4.4}                                                     & $\textrm{INT}_{th}$ (dB)  & -10.5   \\ \hline
polarization loss                                                 & $A_{\textrm{pol}}$ (dB) & 3     \\ \hline
\gls{BS} loading factor \cite{annex4.4}                                                & $\rho$ & 20\%    \\ \hline
\gls{TDD} activity factor \cite{annex4.4}                                                & $F_T$ & 75\%     \\ \hline
Ratio of urban area type \cite{annex4.4}                                          & $R_a$ & 5\%, 10\%  \\ \hline
Ratio of built-up areas \cite{annex4.4}                                          & $R_b$ & 1\%  \\ \hline
\end{tabular}
\end{table}
\subsection{Aggregated interference from the city of Milan}\label{sec:NumericMilan}
\begin{figure}[tb!]
\hspace{-1cm}
\centering
\includegraphics[width = 0.9 \columnwidth]{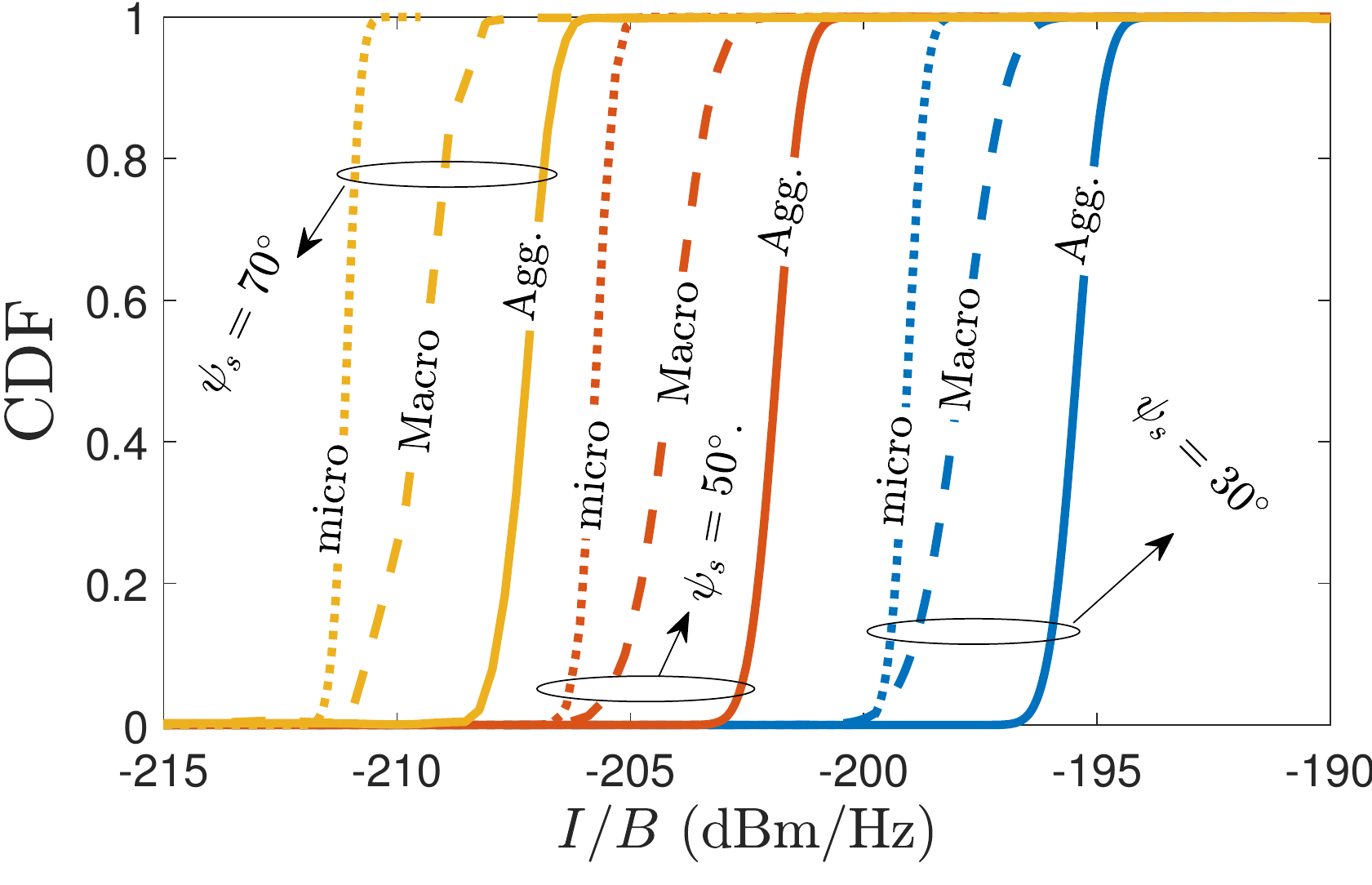}
  \caption{\label{fig:Milan_1} \gls{CDF} of the interference \gls{PSD} using SMI method for the city of Milan, with array configurations 1 of Table \ref{tab:ArrayConf}, and exemplary \gls{SAT} gain $G_s = 20$ dB.}
\end{figure}

Fig. \ref{fig:Milan_1} shows the \gls{CDF} of the aggregated interference power density, using the SMI method with array configuration 1 (Table \ref{tab:ArrayConf}). Given the area of the city of Milan $S_{Milan} = 181.76 \, \textrm{km}^2$, the equivalent number of \gls{BS}s with maximum power is $Q = 155.5$. It can be observed that the INR based on the aggregated interference coming from a city of Milan size is much lower than $\textrm{INR}_{th} = -10.5$ dB. 
%
\subsection{Aggregated interference of \gls{SATFP}}\label{sec:NumericFP}
In order to encompass the interference from the \gls{SATFP}, we follow the methodology introduced in Sec. \ref{Sec:AgglargeArea}. The corresponding specifications of each GC are shown in Table \ref{tab:SATFP_GCs}. The values of $R_a = 5\%$ and $R_a = 10\%$ are chosen according to \cite{Liaison} and \cite{annex4.4}, respectively. The average loading factor $\rho = 20 \%$ corresponds to typical values of coexistence studies when the area under study is a large region consisting of hundreds of BSs or more \cite{annex4.4}.

\begin{figure}[!tb]
    \hspace{-1cm}
    \begin{center}
    \hspace{-1cm}
    \subfloat[][]{\includegraphics[width=1\columnwidth]{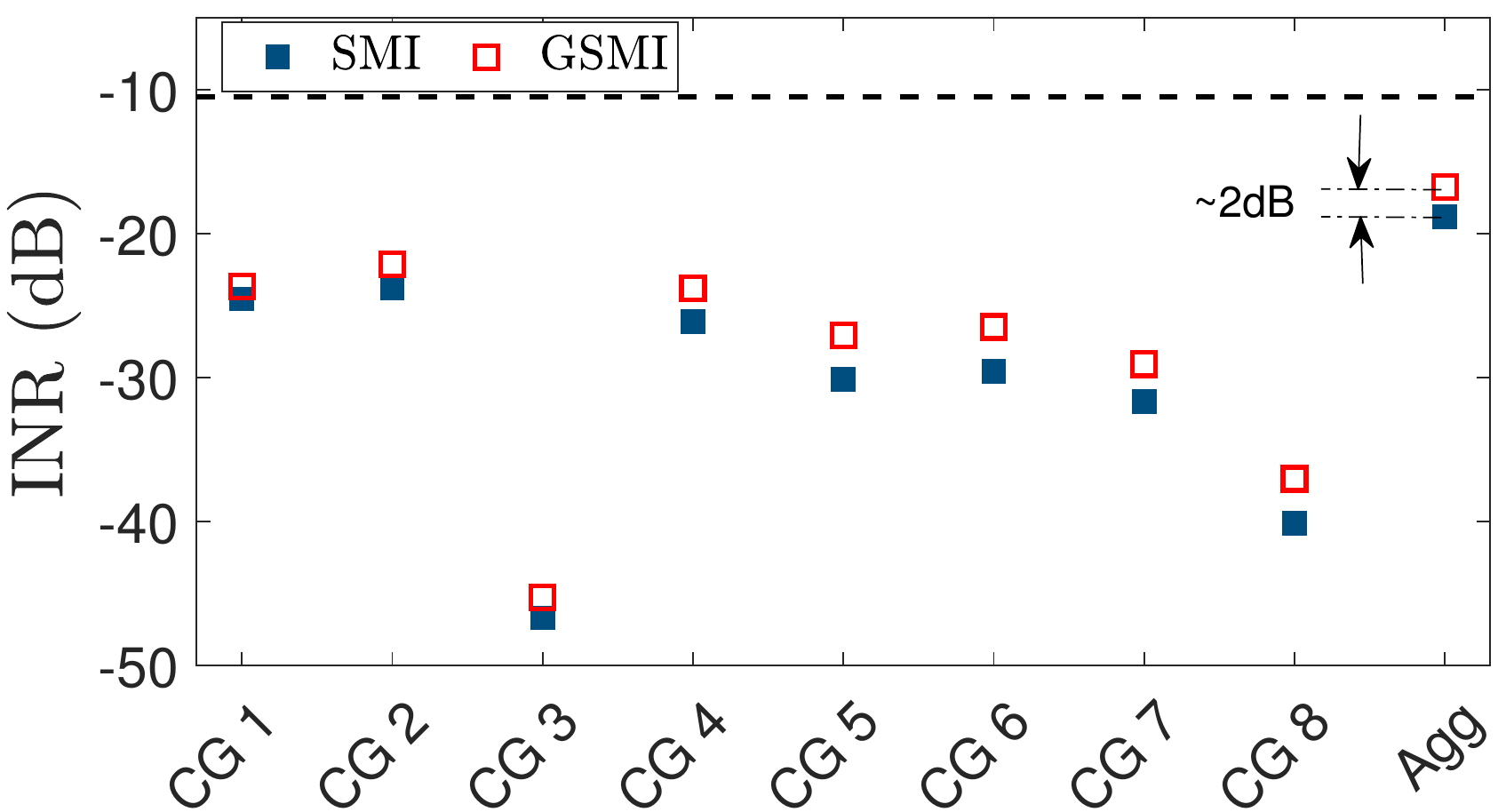}\label{fig:large_region_a}}
    \end{center}
    \begin{center}
    \hspace{-1cm}
    \subfloat[][]{\includegraphics[width=1\columnwidth]{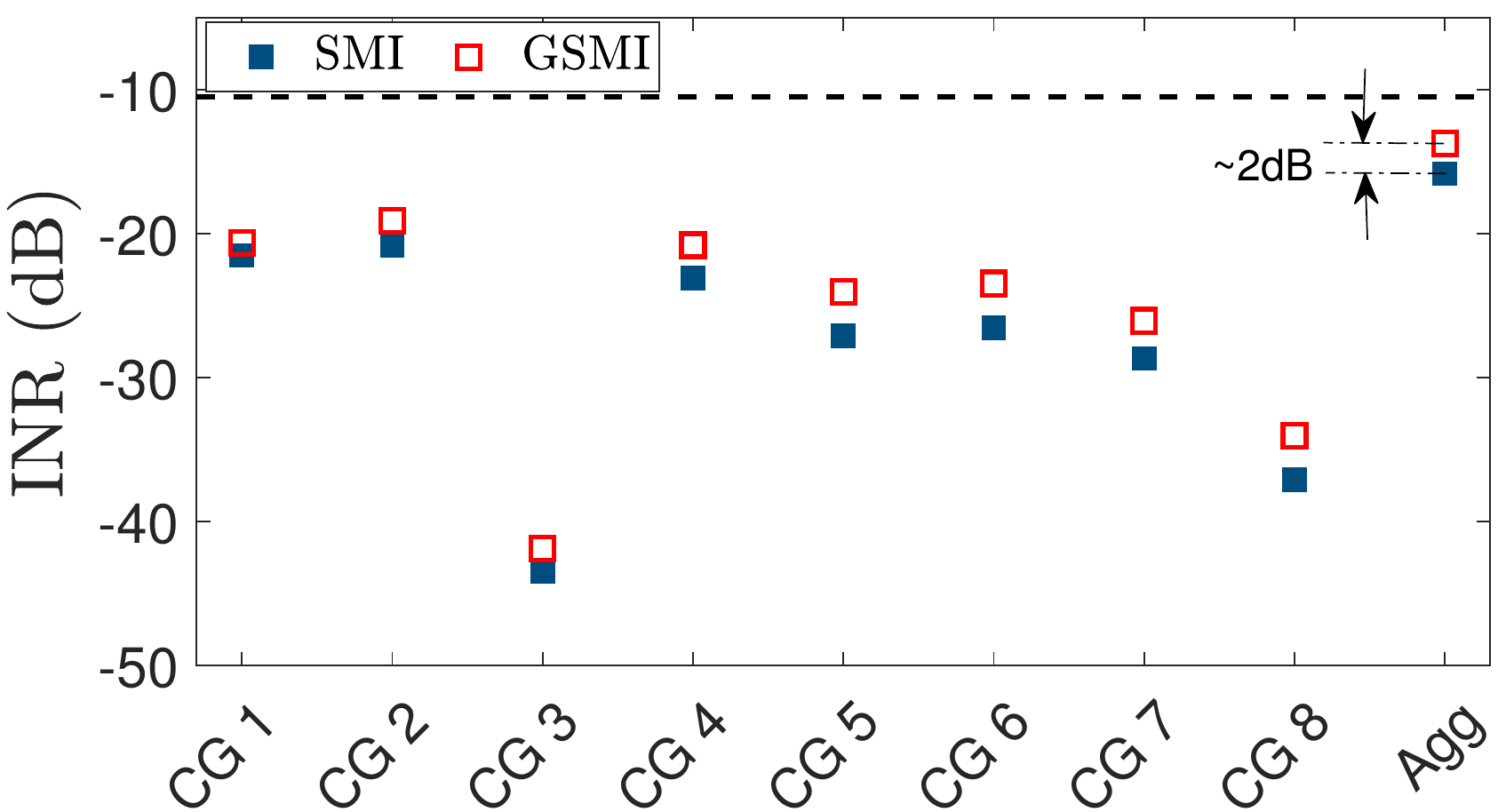}\label{fig:large_region_b}}
    \end{center}
    \caption{\gls{INR} \textcolor{black}{at $80$} percentile for different GCs and the aggregated of \gls{SATFP}, for: a) $R_a = 5\%$; b) $R_a = 10 \%$ (array configuration 1).\label{fig:large_region}
    }
\end{figure}

\begin{figure}[!tb]
\hspace{-1cm}
    \centering
    \includegraphics[width=1\columnwidth]{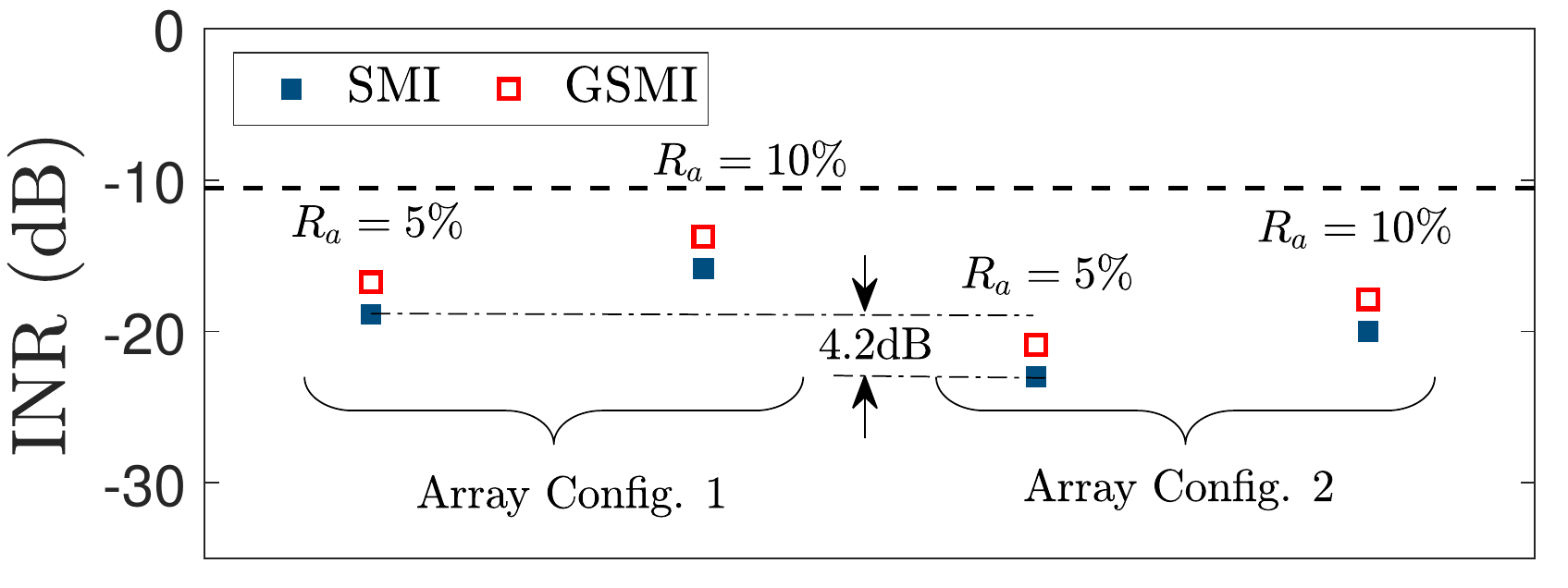}
    \caption{\label{fig:CompareArrayConfigs} \gls{INR} of SATFP \textcolor{black}{at $80$} percentile, comparing the first and second set of \gls{BS} array configurations. \label{fig:Config_Comparison}}
\end{figure}

Fig.~\ref{fig:large_region_a} and \ref{fig:large_region_b} show the median value of the \gls{INR} using the array configuration 1, for 8 GCs and the whole \gls{SATFP} (see Section \ref{Sec:AgglargeArea}), with SMI and GSMI methods. \textcolor{black}{It can be seen that with $Ra = 5\%$, the INR is well below the threshold $\textrm{INR}_{th}$, while only for extremely dense deployments with $R_a = 10\%$, would yield an INR level close to $\textrm{INR}_{th}$. Notice that the SMI method is the baseline model and GSMI is an approximate method that overestimates the interference w.r.t. the SMI by approx. 2 dB.} 
\begin{table}[tb!]
\centering
\caption{Specifications of footprint GCs\label{tab:SATFP_GCs}}
\renewcommand{\arraystretch}{1.2}
\begin{tabular}{|c|c|c|c|}
\hline
$\#GC \, (\upsilon)$ & $G_s^{\upsilon} \,(dBi)$ & $S_{\upsilon} \,(km^{2})$ & $\psi_s^{\upsilon}\, (deg)$ \\ \hline
1                    & 20                       & 3 812 552                 & 30                          \\ \hline
2                    & 20                       & 6 654 033                 & 40                          \\ \hline
3                    & 21                       & 30 088                    & 40                          \\ \hline
4                    & 21                       & 9 203 759                 & 50                          \\ \hline
5                    & 21                       & 5 104 969                 & 60                          \\ \hline
6                    & 22                       & 4 632 108                 & 60                          \\ \hline
7                    & 22                       & 6 869 836                 & 70                          \\ \hline
8                    & 22                       & 2 605 246                 & 80                          \\ \hline
\end{tabular}
\end{table}
One way to reduce the aggregated interference power is to increase the number of array antennas on the vertical plane while keeping constant \gls{EIRP} and preserving the quality of service of the \gls{U6G} service. The consequent reduction of sidelobe's level diminishes the interference at SAT. Fig.~\ref{fig:CompareArrayConfigs} shows the INR for the two array configurations in Table \ref{tab:ArrayConf}. The $80$ percentile of the INR is decreased by more than 4 dB when using the antenna array configuration 2, i.e., double the antennas on the vertical plane of the macro BS. \textcolor{black}{In this latter case, even the extremely dense deployments with $R_a = 10\%$ would be well below the $\textrm{INR}_{th}$.} Other solutions to reduce \gls{U6G} interference reduction can be investigated, but it is beyond the scope of this paper \cite{interf_mitigation}.
%
%
We remark that the numerical results are herein obtained using the statistics of the city of Milan since the statistics of each GC are not available. A more accurate estimation of the level of interference requires accurate knowledge of the distribution of geometric parameters for every different region.
\section{Conclusion}\label{Sec:Conclusion}
In this paper we develop a stochastic model of interference (SMI) to evaluate the aggregated interference power at the SAT in U6G band from a set of \gls{BS}s belonging to an arbitrarily large geographical area. The SMI is based on stochastic array gain and clutter loss, and it considers different \textcolor{black}{interference modes} such as direct path and reflections from buildings and ground. In addition, we propose a geometry-based stochastic model of interference (GSMI) method to be used in the absence of the distribution of diffraction loss and/or reflection loss. We demonstrate, for typical parameters’ values in the context of communications coexistence, that the interference power generated by U6G BSs in typical cases is below the interference thresholds set as tolerable for \gls{SAT} by standardization organizations. Remarkable degrees of freedom for SAT interference reduction is based on how the antenna array and system are designed.
\section*{Acknowledgement}
The research has been carried out in the framework of the Joint Lab between Huawei and Politecnico di Milano. The authors would like to acknowledge the enlightening discussions and clarifications with Carlo Riva on clutter loss models. 
\begin{appendices}
\section{Calculation of occurrence probabilities }\label{Appendix:OccurrenceProb}
\begin{figure}[tb!]
    \centering
    \includegraphics[width = 0.9\columnwidth]{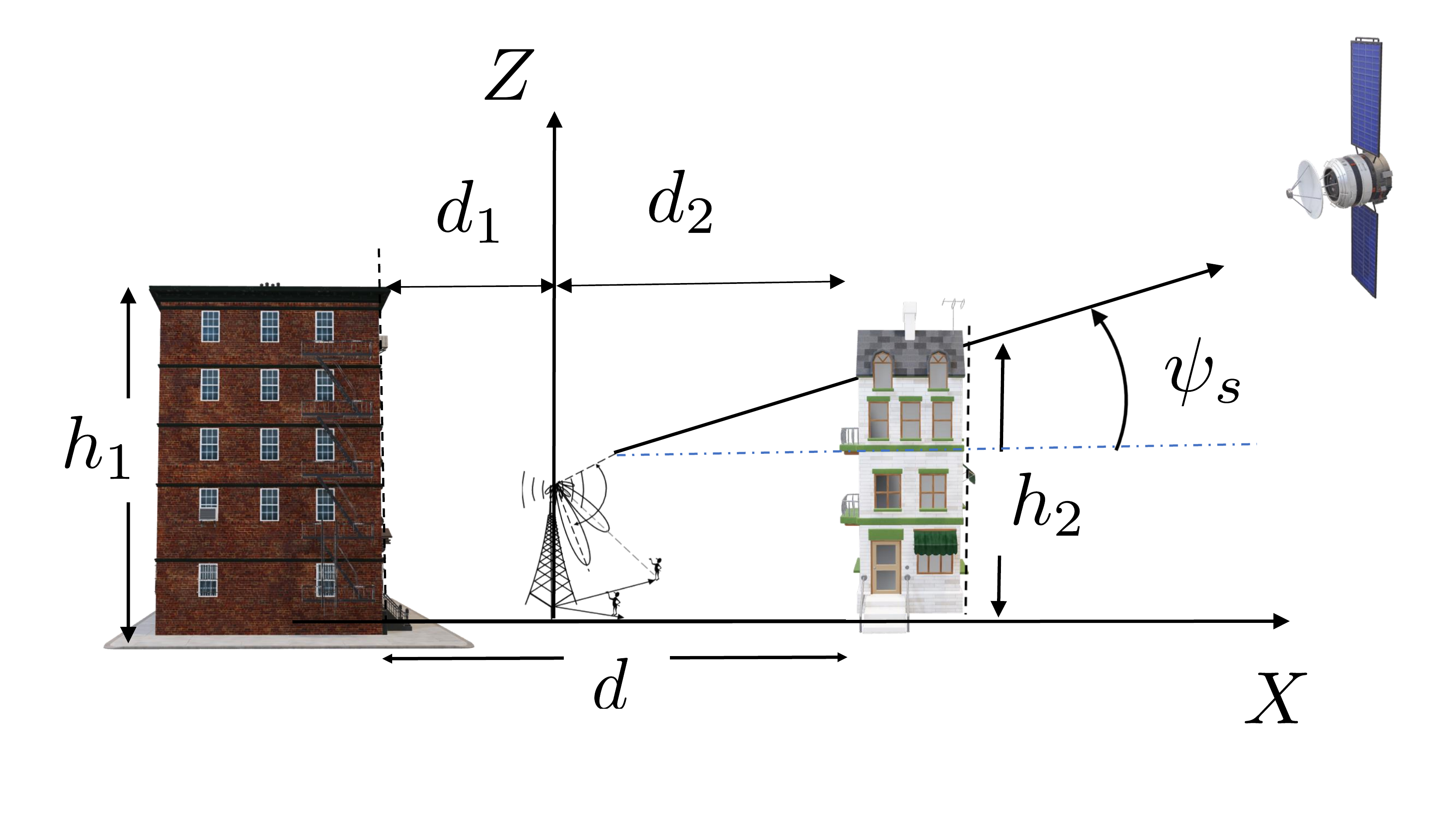}
    \caption{Side view of the direct path propagation mode}
    \label{fig:DP}
\end{figure}

The occurrence probability of the direct path between BS and SAT can be computed from basic geometrical considerations. Based on Fig.~\ref{fig:DP}, the direct path exists whenever it is not blocked by building 2, thus when $h_2 < h_{BS} + d_2\tan(\psi_s)$. Given the \gls{CDF} of the buildings' height, defined as 
\begin{equation*}
    F_h(h) = \int_{0}^h \mathcal{P}(\xi|\phi_s) d\xi,
\end{equation*}
 where $\mathcal{P}(\xi|\phi_s)$ is detailed in Sec.~\ref{Sec:geom_stat}. The occurrence probability of this \textcolor{black}{interference mode} is 
\begin{equation*}
    \mathcal{P}_{DP} = F_h(h_{BS}+d_2\tan(\psi_s)|\phi_s).
\end{equation*} 
Other \textcolor{black}{interference modes} are similarly treated, with straightforward modifications, using the image method (see e.g., \cite{RayTracing1,RayTracing3}). The corresponding occurrence probabilities are not reported for brevity. 
\end{appendices}
\bibliographystyle{IEEEtran}
\bibliography{biblio}

\end{document}